\documentclass[a4paper,fleqn,usenatbib]{mnras}

\usepackage{newtxtext,newtxmath}

\usepackage[T1]{fontenc}
\usepackage{ae,aecompl}
\usepackage{threeparttable,booktabs}
\usepackage{pdflscape} 

\usepackage{graphicx}	
\usepackage{amsmath}	
\title[GLACE survey: galaxy activity in ZwCl0024+1652 cluster]{GLACE survey: galaxy activity in ZwCl0024+1652 cluster from strong optical emission lines}

\author[Z. Beyoro-Amado et al.]{
Zeleke Beyoro-Amado,$^{1,2,3}$\thanks{E-mail: zbamado@gmail.com}
Miguel S\'anchez-Portal,$^{4,5}$
\'Angel Bongiovanni,$^{4,5}$ \newauthor
Mirjana Povi\'c,$^{1,6}$  
Solomon B. Tessema,$^{1}$ 
Ricardo P\'erez-Mart\'inez,$^{5,7}$ \newauthor
Ana Mar\'ia P\'erez Garc\'ia,$^{5,8}$
Miguel Cervi\~no,$^{8}$
Jakub Nadolny,$^{9,10}$\newauthor
Jordi Cepa,$^{9,10}$ 
J. Ignacio Gonz\'alez-Serrano$^{11}$ 
 and
Irene Pintos-Castro$^{12}$ \\
$^{1}$Ethiopian Space Science and Technology Institute (ESSTI), Entoto Observatory and Research Centre (EORC),\\
\ \ Astronomy and Astrophysics Research and Development Division, P.O.Box 33679, Addis Ababa, Ethiopia\\
$^{2}$Kotebe Metropolitan University (KMU), College of Natural and Computational Sciences, Department of Physics, \\
\ \ P.O.Box 31248, Addis Ababa, Ethiopia\\
$^{3}$Addis Ababa University (AAU), P.O.Box 1176, Addis Ababa, Ethiopia\\
$^{4}$Instituto de Radioastronom\'ia Milim\'etrica (IRAM), Av. Divina Pastora 7, N\'ucleo Central, E-18012 Granada, Spain\\
$^{5}$Asociaci\'on Astrof\'isica para la Promoci\'on de la Investigaci\'on, Instrumentaci\'on y su Desarrollo, ASPID, \\
\ \ E-38205 La Laguna, Tenerife, Spain\\
$^{6}$Instituto de Astrof\'isica de Andaluc\'ia (IAA-CSIC),Glorieta de la Astronomia s/n, 18008, Granada, Spain\\
$^{7}$ISDEFE for ESA. Camino Bajo del Castillo s/n. Urb. Villafranca del Castillo. E-28692, Villanueva de la Ca\~nada, Spain\\
$^{8}$Centro de Astrobiolog\'{i}a (CSIC/INTA), ESAC Campus, Camino Bajo del Castillo s/n, E-28692, \\ 
\ \ Villanueva de la Ca\~nada, Spain\\
$^{9}$Instituto de Astrof\'isica de Canarias (IAC), E-38200 La Laguna, Tenerife, Spain\\
$^{10}$Departamento de Astrof\'isica, Universidad de La Laguna (ULL), E-38205 La Laguna, Tenerife, Spain\\
$^{11}$Instituto de F\'{i}sica de Cantabria (CSIC - Universidad de Cantabria), Avda. de los Castros s/n, E-39005, Santander, Spain\\
$^{12}$Department of Astronomy \& Astrophysics, University of Toronto, 50 St. George Street, Toronto, ON M5S 3H4, Canada
}

\date{Accepted 2020 December 8. Received 2020 November 26; in original form 2020 July 23}

\pubyear{2020}

\begin{document}
\label{firstpage}
\pagerange{\pageref{firstpage}--\pageref{lastpage}}
\maketitle

\begin{abstract}
Although ZwCl0024+1652 galaxy cluster at $z\sim0.4$ has been thoroughly analysed, it lacks a comprehensive study of star formation and nuclear activity of its members. With GaLAxy Cluster Evolution (GLACE) survey, a total of 174 H$\alpha$ emission-line galaxies (ELGs) were detected, most of them having [N{\sc ii}]. We reduced and analysed a set of [O{\sc iii}] and H$\beta$ tunable filter (TF) observations within GLACE survey. Using H$\alpha$ priors, we identified [O{\sc iii}] and H$\beta$ in 35 ($\sim$20\%) and 59 ($\sim$34\%) sources, respectively, with 21 of them having both emission lines, and 20 having in addition [N{\sc ii}]. Applying BPT-NII diagnostic diagram, we classified these ELGs into 40\% star-forming (SF), 55\% composites, and 5\% LINERs. Star formation rate (SFR) measured through extinction corrected H$\alpha$ fluxes increases with stellar mass ($\mathrm{M}_{*}$), attaining its peak at $\mathrm{M}_{*}\sim10^{9.8}\mathrm{M}_\odot$. We observed that the cluster centre to $\sim$\,1.3\,Mpc is devoid of SF galaxies and AGN. Our results suggest that the star formation efficiency declines as the local density increases in the cluster medium. Moreover, the SF and AGN fractions drop sharply towards high-density environments. We observed a strong decline in SF fraction in high $\mathrm{M}_*$, confirming that star formation is highly suppressed in high-mass cluster galaxies. Finally, we determined that SFR correlates with $\mathrm{M}_*$ while specific SFR (sSFR) anti-correlates with $\mathrm{M}_*$, both for cluster and field. This work shows the importance and strength of TF observations when studying ELGs in clusters at higher redshifts. We provide with this paper a catalogue of ELGs with H$\beta$ and/or [O{\sc iii}] lines in ZwCl0024+1652 cluster.
\end{abstract}



\section{Introduction}
\label{sec:intro}
Galaxy clusters are some of the largest known gravitationally bound structures, consisting of hundreds to thousands of galaxies \citep[e.g.][]{Press1974,Trenti2012}. Investigating the way galaxies form and evolve in both field and clusters gives a very important information for the concern of modern cosmology. Moreover, to understand how galaxies transform inside the clusters presents one of the main steps in disentangling the picture of galaxy formation and evolution, and the formation and evolution of the universe at large. Some of previous works observed differences between the properties of galaxies in the field and those in clusters \citep[e.g.,][]{Dre85,Bal97,Koo98,Cucciati2010,Chung2011,Hwang2012,Delaye2014,Cantale2016,Allen2016,Maier2016,Saracco2017,Matharu2020,Old2020,Perez-Martinez2020,Tiley2020}, although the difference seems to be more significant for star-forming than for quiescent galaxies \citep[e.g.,][]{Allen2016,Saracco2017,Old2020,Tiley2020}. Specifically,  the star formation activity seems to be anti-correlated with the local density \citep{Haines2013,Kovac2014}. Considering the effect of global environment, it is observed that the fraction of blue galaxies is higher in field than in clusters up to redshift of $\sim$\,1.2 \citep{Muzzin2012}. Concerning the variations with local density, the fractions of SF and blue galaxies seem to decline as the local density increases, in low- to intermediate-redshift clusters \citep[e.g.,][]{Kodama2004,Baldry2006,Cooper2008,Cucciati2017}. On the other hand, there are studies showing correlation of the rate of quenching and the fraction of quiescent galaxies in relation to the stellar mass and local environment \citep[e.g.,][]{Baldry2006,Vulcani2010,Kovac2014,Darvish2017,Nantais2017}. Moreover, a well established relation called morphology--density relation is in place \citep{Dressler1980}, showing that the fraction of early--type (ET) galaxies increases with local density while the fraction of late--type (LT) galaxies decreases.

Moreover, significant evolution in the properties of galaxies in clusters have been observed as a function of redshift as well as environment \citep[e.g.,][]{Alt10,Coi05,Gea06, Moran2007,Hai09, Koy11,Mar13,PiC13,Sanchez2015,PC2016}. The study of AGN population, star formation (SF) activity, morphology, colour-magnitude relations, and distribution of galaxy metallicities with cluster properties (e.g., cluster radii and local density) present the powerful mean to investigate evolution within clusters. Concerning morphology being one of the important galaxy properties, the cores of nearby clusters are dominated by red ET galaxies, while at higher redshifts the population of blue-dominated LT galaxies increases \citep[e.g.,][]{BOe84,Bow98,PC2016}, the so called BO effect. In \citet{Amado2019}, we performed the morphological classification of galaxies in ZwCl\,0024\,+\,1652 cluster \citep[applying the same method as in][]{Povic2012,Povic2013,Povic2015,PC2016} using HST/ACS data confirming that the core of this cluster is dominated by ET galaxies over the LT ones. Regarding SF activity, the increase of the obscured SF was observed in mid and far infrared (IR) surveys of distant clusters \citep[e.g.,][]{Coi05,Gea06,Hai09}. 

Different results have been reported regarding AGN activity in clusters. However, the way how the AGN fraction varies with environment, and the role of AGN in galaxy transformations and evolution in galaxy clusters is still not clear. Some studies suggest that AGN activity increases with redshift \citep[e.g.,][]{Mar02,Eas07,Siv08,Mar13} as revealed using X-ray data. There are studies that suggest that AGN fraction does not depend on the local density at all \citep[e.g.,][]{Miller2003,Linden2010,Pimbblet2013,PC2016,Marziani2017} whereas others argue that high density regions and luminous AGNs look disjoint \citep[e.g.,][]{Kauffmann2004,Alonso2007,Argudo-Fernandez2018,Bornancini2020}. 

Following the hierarchical model of structure formation galaxies merge into larger systems with time. This process is likely responsible for the evolution of the proprieties of cluster galaxies, as a function of redshift and environment \citep[e.g.,][]{Bal00,KoB01}. Different physical processes like: mergers and harassment (both resulting from galaxy-galaxy low- and high-speed interactions, respectively), starvation (slow decrease in SF), ram-pressure stripping, thermal interstellar mass evaporation, turbulent stripping, etc. were suggested for affecting the galaxy evolution in clusters \citep[e.g.,][]{GunnGott1972,Larson1980,Moore1996,Cole2000,Nipoti2007,vandenBosch2008,Roediger2009,Dressler2013,Dzudzar2019,Coccato2020}. These processes may result from galaxy--inter-cluster medium (ICM) interactions, tidal triggering SF, or tidal halo stripping, depending on the distance from the cluster core \citep{Treu2003}. These physical processes act on the emission line galaxies (ELG) population (both SF and AGN) of the cluster \citep[e.g.,][]{Poggianti2017}. Apart from these intra-cluster physical processes, \citep[e.g.,][]{Sobral15} suggest that rapid and significant star formation and AGN activity could be addressed to massive cluster merging. In identifying the cluster ELGs, narrow-band imaging surveys have shown to be very efficient \citep{Koy2010,Sanchez2015}. The central regions of clusters are devoid of ELGs (SF and AGN) such as H$\alpha$ and mid-IR emitters, and the SFRs of clusters decrease rapidly since $z\,\sim$\,1 \citep[e.g.,][]{Koy2010,Koulouridis2019}.

GaLAxy Cluster Evolution (GLACE) survey has been designed by taking advantage of the power of narrow-band imaging. It is an innovative survey of ELGs and aims in studying SF and AGN activity, morphology, and metallicity variations of galaxies in clusters at different redshifts \citep{PiC13,Sanchez2015,PC2016,Amado2019}. The sample of clusters at three redshifts ($z\,\sim$\, 0.40, 0.63, and 0.86) have been selected. The mapping of the strongest optical emission lines such as H$\beta$, H$\alpha$, [N{\sc ii}]6583, [O{\sc ii}]3727, and [O{\sc iii}]5007 have been carried out in the selected spectra windows scanned by the TF observation. These spectral windows were chosen to be relatively free from the strong OH emission lines \citep{Rousselot2000}. 

In this paper, we use the GLACE narrow band tunable filter (TF) imaging data of an intermediate redshift cluster ZwCl0024+1652 at $z\,=\,0.395$. This cluster exhibit the so called BO effect \citep{BOe84}. The core of the cluster is dominated by red ET galaxies \citep[e.g.,][and references therein]{Amado2019}, with slightly increasing LT galaxies outwards. Moreover, \citet{Czoske2001,Czoske2002} identified substructures in foreground and background with wide field spectroscopic analysis. They found bimodal redshift distribution which suggests that the cluster is composed of two systems colliding (merging) along the line of sight. Even though the cluster and its member galaxies have been analysed thoroughly by different authors \citep[e.g.,][]{Kneib2003,Treu2003,Moran2007,Geach2009,Natarajan2009,Sanchez2015,Amado2019}, it still lacks a comprehensive study of SF and AGN properties of its member galaxies. In this work we focus on analysing the [O{\sc iii}] and H$\beta$  emission lines for improving our knowledge on SF and AGN activity obtained through previous analysis of H$\alpha$  and [N{\sc ii}] emission lines \citep{Sanchez2015}.

The paper is organized as follows: Section~\ref{sec:data} describes [O{\sc iii}] and H$\beta$ TF data used in this work; in Section~\ref{sec:Reduction}, general data reduction has been described. Determination of the list of emission lines in [O{\sc iii}] and H$\beta$ has been done in Section~\ref{sec:Emitters}. We present the results and analysis in Section~\ref{sec:Results}. We discussed our results in Section~\ref{sec:Discussions} and finally, conclusions are summarized in Section~\ref{sec:Conclusions}.

The following cosmological parameters are assumed throughout this paper: $\Omega_M=0.3$, $\Omega_\Lambda=0.7$, $\Omega_k=0$ and $H_0=70\,Km\,s^{-1}\,Mpc^{-1}$. All magnitudes are given in AB system as described by \citet{OkeGun1983}, unless otherwise stated.

\section{Data}
\label{sec:data}
ZwCl0024+1652 is a rich galaxy cluster at redshift z\,$=$\,0.395, notable due to evidence of a possible dark matter ring around the cluster core \citep{Jee2007}. \citet{Kodama2004} have identified 2385 cluster members by computing the photometric redshift of galaxies brighter than $z'=22.7$ over a total area of 712.79\,\textrm{arcmin}$^2$. The brightest cluster galaxy (BCG) has r--band magnitude of $\sim$\,17.74 \citep{Wen2013}. As determined by \citet{Kneib2003} with lensing analysis, the cluster viral radius is R$_{200}$\,=\,1.7\,\textrm{Mpc} with a total mass of $\sim$\,5.7\,$\times$\,10$^{14}M_\odot$. The cluster was determined to be characterized by at least two dynamical structures as described in Section~\ref{sec:Results}. As reported by \citet{Sanchez2015} using the GLACE data, the H$\alpha$ luminosity function is well aligned with estimates from \citet{Kodama2004}. In the following, we describe the data used in this work for carrying out the further studies of ZwCl0024+1652 cluster. 
\subsection{The GLACE TF data}
\label{sec:GLACE_data}
Two pointings have been carried out with OSIRIS/GTC using the red TF towards Zwcl0024+1652 targeting the [O{\sc iii}] and H$\beta$ lines. The first pointing (centred at RA\,=\,00$^{\rm h}$\,26$^{\rm m}$\,35.70$^{\rm s}$, DEC\,=\,17$^{\rm d}$\,09$^{\rm m}$\,45.0$^{\rm s}$) referred as ``centre position'' or ``position A'' was executed in GTC semesters 8-10AGOS, 75-13B and 63-09B. In the centre position observations the core of the cluster was centred within CCD1\footnote{The mosaic of OSIRIS detector is composed of two 2048\,$\times$\,4096 pixel CCDs separated, with a plate scale of 0.125 arcsec/pixel. Refer to \citet{Cepa2005} as well as the OSIRIS Users' manual at \tt http://www.gtc.iac.es/instruments/osiris/media.}. The second pointing observations (centred at RA\,=\,00$^{\rm h}$\,26$^{\rm m}$\,26.5$^{\rm s}$, DEC\,=\,17$^{\rm d}$\,12$^{\rm m}$\,15.0$^{\rm s}$) referred as ``offset position'' or ``position B'' were performed in GTC semesters 75-13B and 47-10B. The position B offsets are  $\Delta$RA = -2.3 arcmin, $\Delta$DEC = +2.5 arcmin; i.e.,  about 3.4 arcmin in the NW direction. 

The [O{\sc iii}] line observations spanned at the spectral range of 6894.8\,--\,7154.8\,\AA\ that was covered by 27 evenly spaced scan steps of $\Delta\lambda$\,=\,10\,\AA. Hence, there are 27 individual image slices for each position (centre position and offset position) corresponding to each wavelength. In GTC semesters 8-10AGOS and 47-10B, three individual exposures have been carried out for each TF tune with a ``L-shaped'' dithering pattern with an amplitude of  10\,arcsec (similar to the gap between the detectors), implemented to easily identify the diametric ghosts (see \cite{Jones2002} for a detailed description), while in GTC semester 75-13B, only two individual dithered exposures for each TF tune have been mapped for the same wavelength steps. The observational data of [O{\sc iii}] imaging is summarized in Table~\ref{tab:OIII}.

\begin{table*}
	\centering
	\caption{ZwCl0024+1652 imaging data targeting [O{\sc iii}] lines.}
	\label{tab:OIII}
	\begin{tabular}{lllllll}
	 \textbf{Centre position:} [{\rm RA}\,=\,00$^{\rm h}$\,26$^{\rm m}$\,35.70$^{\rm s}$, {\rm DEC}\,=\,17$^{\rm d}$\,09$^{\rm m}$\,45.0$^{\rm s}$] \\
	\end{tabular}
	\begin{tabular}{lllllll} 
	\hline\hline
\textbf{$\lambda_{\rm 0,i}$} & \textbf{OS filter} & \textbf{Date} & \textbf{Seeing} & \textbf{N-Steps} & \textbf{N-Exp.} & \textbf{Exp-time} \\
(\AA) & & & ('') & & & (sec.) \\
		\hline\hline
6894.8 & f680/43 & 28 July 2010 & 0.9\,-\,1.3 & 4 & 3 & 45 \\
6934.8 & f694/44 & 28 July 2010 & 1.0\,-\,1.3 & 14  & 3 & 45 \\
7074.8 & f709/45 & 28 July 2010 & 0.9\,-\,1.3 & 9 & 3 & 45 \\
6894.8 & f680/43 & 09 October 2013 & <\,1.0 & 4 & 2 & 53 \\
6934.8 & f694/44 & 09 October 2013 & <\,1.0 & 14 & 2 & 53 \\
7074.8 & f709/45 & 09 October 2013 & <\,1.0 & 9 & 2 & 53 \\
	\hline
\end{tabular}
	 \begin{tabular}{lllllll}
	 \\
	 \textbf{Offset position:} [{\rm RA}\,=\,00$^{\rm h}$\,26$^{\rm m}$\,26.5$^{\rm s}$, {\rm DEC}\,=\,17$^{\rm d}$\,12$^{\rm m}$\,15.0$^{\rm s}$] \\
	\end{tabular}
	\begin{tabular}{lllllll} 
	\hline\hline
\textbf{$\lambda_{\rm 0,i}$} & \textbf{OS filter} & \textbf{Date} & \textbf{Seeing} & \textbf{N-Steps} & \textbf{N-Exp.} & \textbf{Exp-time} \\
(\AA) & & & ('') & & & (sec.) \\
		\hline\hline
6894.8 & f680/43 & 20 September 2010 & 0.9\,-\,1.3 & 4 & 3 & 45 \\
6934.8 & f694/44 & 20 September 2010 & 1.0\,-\,1.3 & 14  & 3 & 45 \\
7074.8 & f709/45 & 20 September 2010 & 0.9\,-\,1.3 & 9 & 3 & 45 \\
6894.8 & f680/43 & 09 October 2013 & <\,1.0 & 4 & 2 & 53 \\
6934.8 & f694/44 & 09 October 2013 & <\,1.0 & 14 & 2 & 53 \\
7074.8 & f709/45 & 09 October 2013 & <\,1.0 & 9 & 2 & 53 \\
	\hline
	\end{tabular}
	\end{table*}

The spectral range of the H$\beta$ observations was 6701.1\,-\,6911.1\,\AA. For this line, the target was covered by 22 evenly spaced scan steps of $\Delta\lambda$\,=\,10\,\AA\ at the centre position; at the offset position, two scan steps at 6901.1\,\AA\ and 6911.1\,\AA\ were omitted accidentally, hence we only have 20 image slices for position B. As in the previous case, three individual dithered exposures with a ``L-shaped'' pattern were obtained for each TF tune.  The H$\beta$ imaging data are summarised in Table~\ref{tab:HB}.

\begin{table*}
	\centering
	\caption{ZwCl0024+1652 imaging data targeting H$\beta$ lines.} 
	\label{tab:HB}
	\begin{tabular}{lllllll}
	 \textbf{Centre position:} [{\rm RA}\,=\,00$^{\rm h}$\,26$^{\rm m}$\,35.70$^{\rm s}$, {\rm DEC}\,=\,17$^{\rm d}$\,09$^{\rm m}$\,45.0$^{\rm s}$] \\
	\end{tabular}
	\begin{tabular}{lllllll} 
	\hline\hline
 \textbf{$\lambda_{\rm 0,i}$} & \textbf{OS filter} & \textbf{Date} & \textbf{Seeing} & \textbf{N-Steps} & \textbf{N-Exp.} & \textbf{Exp-time} \\
 (\AA) & &  & ('') & & & (sec.) \\	
		\hline\hline
6701.1 & f666/36 & 24 November 2009 & 0.8\,-\,1.0 & 5 & 3 & 124 \\
6751.1 & f666/36 & 25 November 2009 & 1.1\,-\,1.2 & 3 & 3 & 124 \\
6781.1 & f680/43 & 25 November 2009 & 0.8\,-\,1.0 & 2 & 3 & 124 \\
6801.1 & f680/43 & 17 August 2010 & 1.0 & 3 & 3 & 130 \\
6831.1 & f680/43 & 24 November 2009 & 0.9\,-\,1.1 & 5 & 3 & 124 \\
6881.1 & f680/43 & 24 November 2009 & 0.8\,-\,1.0 & 4 & 3 & 124 \\
	\hline
\end{tabular}
	 \begin{tabular}{lllllll}
	 \\
	 \textbf{Offset position:} [{\rm RA}\,=\,00$^{\rm h}$\,26$^{\rm m}$\,26.5$^{\rm s}$, {\rm DEC}\,=\,17$^{\rm d}$\,12$^{\rm m}$\,15.0$^{\rm s}$] \\
	\end{tabular}
	\begin{tabular}{lllllll} 
	\hline\hline
 \textbf{$\lambda_{\rm 0,i}$} & \textbf{OS filter} & \textbf{Date} & \textbf{Seeing} & \textbf{N-Steps} & \textbf{N-Exp.} & \textbf{Exp-time} \\
(\AA) & & & ('') & & &(sec.) \\
 		\hline\hline
6701.1 & f666/36 & 09 October 2013 & <\,1.0 & 5 & 3 & 138 \\
6751.1 & f666/36 & 09 October 2013 & <\,1.0 & 3 & 3 & 138 \\
6781.1 & f680/43 & 09 October 2013 & <\,1.0 & 2 & 3 & 138 \\
6801.1 & f680/43 & 09 October 2013 & <\,1.0 & 5 & 3 & 138 \\
6851.1 & f680/43 & 09 October 2013 & <\,1.0 & 5 & 3 & 138 \\
	\hline
	\end{tabular}
	\end{table*}

\subsection{Ancillary cluster data}
\label{sec:ancillary_data}
In addition to the GLACE data, we used a public master catalogue\footnote{\tt http://www.astro.caltech.edu/$\sim$smm/clusters/} data for ZwCl0024+1652 cluster described in \citet{Treu2003} and \citet{Moran2005} consisting of 73\,318 sources covering the area of 0.5\,$\times$\,0.5\,deg$^2$ up to the clustercentric distance of $\sim$\,5\,Mpc. This catalogue provides the morphological classification of sources brighter than $I\,=\,22.5$ \citep{Moran2005} and photometric data used in our analysis. The morphological classification was performed visually where spiral, lenticular or elliptical morphologies are assigned if clearly resolved. Moreover, for the visual inspection of thumbnails we used the HST/ACS\footnote{Based on observations made with the NASA/ESA HST, and obtained from the Hubble Legacy Archive, which is a collaboration between the Space Telescope Science Institute (STScI/NASA), the Space Telescope European Coordinating Facility (ST-ECF/ESA) and the Canadian Astronomy Data centre (CADC/NRC/CSA)} high resolution reduced scientific images observed using the ACS Wide Field Camera (WFC) having a pixel scale of 0.05 arcsec per pixel and field of view of $202\times202$ arcsec$^2$. We also used morphological classification of galaxies from \citet{Amado2019}. In this work authors used galaxy Support Vector Machine \citep[galSVM;][]{HC2008} to classify cluster member galaxies up to a clustercentric distance of 1\,\textrm{Mpc}.
\subsection{Galaxy data for cluster-field comparison}
\label{sec:field_data}
In order to compare selected properties of field galaxies with the cluster parameters described in this work, we explored different observed spectroscopic data sets which match the flux depth and redshift domains explored by GLACE survey, in particular the surveyed one of ZwCl0024+1652 at z\,$=$\,0.395. We found that the VIMOS-VLT Deep Survey \citep[VVDS-Deep;][]{LeFevre2013} data sample in the field 0226-04 (0.61 deg$^2$) meets this purpose. This publicly available database\footnote{\tt http://cesam.lam.fr/vvds} contains reduced fluxes and equivalent widths (EW) of strong lines in the optical, as well as SFR and $M_*$ estimations of 11\,486 sources limited to $17.5\leq I_{\rm AB}\leq 24$. A detailed description of these measurements can be found in \cite{Lamareille2009}. According to these authors, VVDS-Deep data has a mass-to-light completeness of 50\% at z\,$\sim$\,0.4 and $\log (M_*/M_\odot) \geq 8.5$. With this mass constraint, we prepared a representative field galaxy sample by selecting from VVDS-Deep data all sources within a redshift range (0.37 $\leq$ z $\leq$ 0.42) with a continuum flux ($R_{\rm AB} < 23$), according to the cluster galaxy distributions given by \cite{Sanchez2015}. A total of 177 galaxies satisfy these criteria, increasing the final completeness of the sample to about 90\%.
\section{Data reduction and Calibration}
\label{sec:Reduction}
We carried out the basic data reduction using a TF data reduction ({\tt TFReD}) package developed by \citet{Jones2002} and modified by GLACE members to fit OSIRIS data \citep{Sanchez2015,RamonPerez2019a,RamonPerez2019b,Bongiovanni2019,Bongiovanni2020}. In addition, the {\tt IDL} and {\tt Python} scripts tailored for our specific data sets, and {\tt IRAF}\footnote{\tt http://ast.noao.edu/data/software} have been used. 

\begin{figure}
    \centering
    \includegraphics[width=\columnwidth]{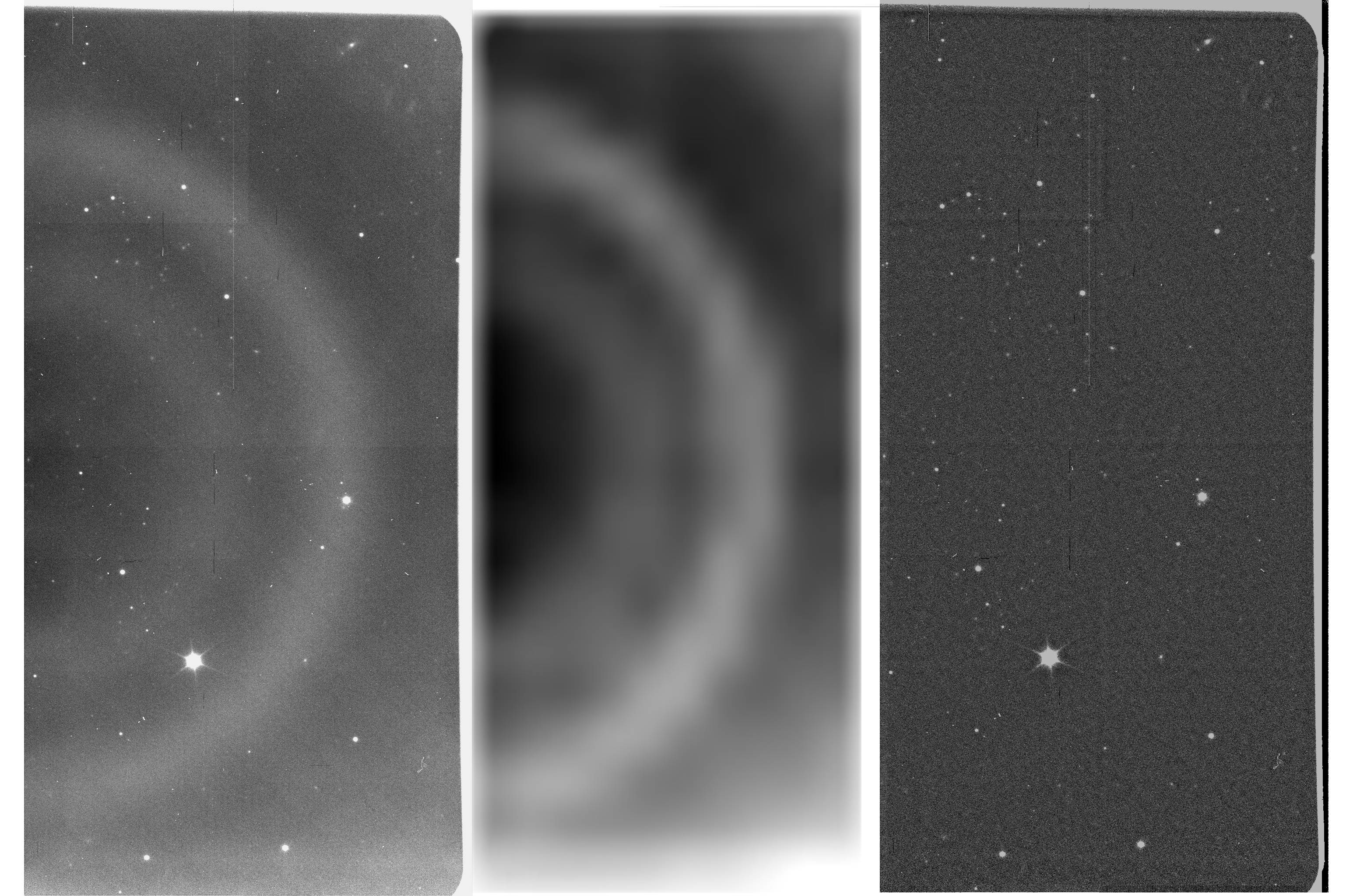}
    \caption{Example of image of H$\beta$ obtained for removing the sky rings using {\tt TFReD} package. This is a CCD2 image obtained with the TF tuned at 6881.1\,\AA\, before subtraction of the sky rings (left), the background sky map created by the {\tt tringSub2} task (centre), and the same image after the subtraction of the sky rings (right).}
    \label{fig:sky_sub1}
\end{figure}

For basic steps of data reduction related with bias, flat-field, and cosmic rays rejection standard {\tt IRAF} procedures have been followed. We started the tunable filter data reduction procedures by removing the sky rings resulting from a radially changing wavelength shift as described by \citet{Gonzalez2014}, following the other steps of {\tt TFReD} data reduction as has been done in \citet{Sanchez2015}. Sky ring corrected images have been prepared for source detection, as can be seen in Figure~\ref{fig:sky_sub1}.

The TF tomography technique in GLACE comprises of scans carried out with dithering described by \citet{LaraLopez2010}. Hence we performed combinations with median filter \citep{Sanchez2015} and created a final deep image in each case ([O{\sc iii}] and H$\beta$) considering the dithering. We extracted the sources from each deep image with {\tt SExtractor} \citep{BerArn1996}.

Considering two images (image1 and image2) with the same central wavelength ($\lambda$) in two different dithering positions, the wavelength difference for a particular source (at respective distances of r$_1$ and r$_2$ in arcsec to the optical centre) was estimated as: 
\begin{equation}
\begin{split}
 \Delta\lambda = \lambda_1-\lambda_2 = 5.04({r_2}^2-{r_1}^2)+(6.0396-1.5698 \\ \times10^{-3}\lambda+1.0024\times10^{-7}\lambda^2)\times({r_1}^3-{r_2}^3)
 \label{eq:dwavel}
 \end{split}
\end{equation}
For tuned wavelength and each detected source we determined the best possible combination and derived the equivalent transmission profile. This resulted with centre/offset possible combinations of 1301/1162 and 325/298 in the case of [O{\sc iii}] and H$\beta$, respectively. We selected a catalogue of 266 sources from GAIA \citep{Lindegren2016,Lindegren2018} covering the entire field with sky positions of reference objects for mapping the cluster by performing astrometry and then determined the astrometric solution with standard {\tt IRAF} tasks {\tt ccxymatch} and {\tt ccmap}. To measure the flux at each slice and source position, we generated set of degraded images using {\tt TFReD} specific task; {\tt tfwhm} (for measuring the FWHM) and {\tt tgauss}. For more detailed description see \citet{Sanchez2015}.

Having a catalogue from deep image and performing photometry on the degraded images, we generated a raw catalogue of candidates by using {\tt tsex} and {\tt tespect} routines of {\tt TFReD}. We separated the non-star-like from from star-like sources and produced flux and flux-error files, containing a non calibrated pseudo-spectra and a corresponding error in flux, respectively. For [O{\sc iii}] and H$\beta$ centre/offset position we detected a total of 724/1168 and 945/1277 non-star-like candidates, respectively. A pseudo-spectrum is a file consisting of tuples with wavelength ($\lambda$) at source position and flux value. It differs from a standard spectrum produced by dispersive medium in that it's flux at each point results from integrating within a passband of the filter centred at the point's wavelength as depicted in \citet{Sanchez2015}. 

For wavelength calibration we first determined the optical centre and the effective wavelength of the observation for each image. Based on \citet{Sanchez2015}, we considered the solution to converge for consecutive wavelength difference lower than 1\,\AA. We followed a two step procedure in performing the flux calibration of each TF tune. The first step is to derive the total efficiency $\epsilon(\lambda)$ of the system (telescope, optics and detector) using a flux measured with aperture photometry and a published flux at given wavelength for set of exposures of the spectrophotometric standard stars (Table~\ref{tab:spectro}). 

\begin{table}
	\centering
	\caption{Spectrophotometric standard stars; the top three for centre while the bottom two for offset position observations.}
	\label{tab:spectro}
	\begin{tabular}{lll} 
		\hline
	        \hline
Name & Mag ($\lambda,$ \AA) & Reference \\
		\hline\hline
G158-100 & 14.87 (5400) & \citet{Filippenko1984} \\
L1363-3 & 13.28 (5560) & \citet{Oke1974} \\
Feige 110 & 11.82 (5556) & \citet{Massey1990} \\
G191-B2B & 11.72 (5460) & \citet {Oke1990} \\ 
G157-34 & 15.35 (5400) & \citet{Filippenko1984} \\ 
	\hline
\end{tabular}
\end{table}
The final step in flux calibration is converting the extracted flux to physical units. For each source at each single i-${th}$ exposure, having the total efficiency $\epsilon(\lambda)$ already estimated, we converted the measured flux $f_{\rm ADU}(\lambda)_i$ to actual flux $f_{\rm m}(\lambda)_{\rm i}$ in $\mathrm{erg}\,\mathrm{~s}^{-1}\,\mathrm{cm}^{-2}$\,\AA$^{-1}$ by using:
\begin{equation}
f_m(\lambda)_{\rm i}=f_{ADU}(\lambda)_{\rm i}\frac{g\,K(\lambda)\,E_\gamma(\lambda)}{\epsilon(\lambda)\,t\,A_{\rm tel}\,\delta\lambda_{\rm e}},
\label{eq:convert}
\end{equation}
where g denotes the CCD gain in $\mathrm{e}^{-1}\,\mathrm{ADU}^{-1}$, $\mathrm{E}_\gamma(\lambda)$ stands for the energy of photon in $\mathrm{erg}$, t is the exposure time in sec, $\mathrm{A}_{\rm tel}$ is the area of the primary mirror of the telescope in $\mathrm{cm}^2$, $\delta\lambda_{\rm e}$ is the effective passband width in \AA, while $K(\lambda)$ stands for atmospheric extinction correction. More detailed information on data reduction and calibration can be found in in \citet{Sanchez2015}.

\section{The [O{\sc iii}] and H$\beta$ emitters}
\label{sec:Emitters}
\subsection{Identifying the emission line candidates through H$\alpha$ priors}
\label{sec:priors}
We cross-matched [O{\sc iii}] non-star-like catalogues of 724/1168 centre/offset sources with the catalogue of ELGs confirmed through H$\alpha$  priors \citep{Sanchez2015} using a radius of 1\,arcsec and found 114 and 95 counterparts for centre and offset positions, respectively, with 57 common sources. In \citet{Sanchez2015}, 174 H$\alpha$ emitters with derived redshifts were confirmed after removing contaminants \citep{Kodama2004}. The redshifts are distributed in $0.35\le\,z\,\le\,0.45$ range and were checked to be consistent by comparing with spectroscopic redshifts from \cite{Moran2005}.
After merging results from both positions and removing common sources, we identified a total of 152 unique ELGs as preliminary [O{\sc iii}] emission line candidates with H$\alpha$ counterparts. 

Similarly, in case of H$\beta$ data we obtained 945 and 1277 sources in centre and offset position, respectively. After cross-matching both catalogues with H$\alpha$ ELGs, we obtained 127 and 102 matches for the centre and offset positions, respectively, with 67 duplicates. Finally, we identified a total of 162 ELGs as preliminary candidates for H$\beta$ line emission. For all the [O{\sc iii}] and H$\beta$ candidates we measured their fluxes, as described in following sections. 

\subsection{Computation of the [O{\sc iii}] line fluxes}
\label{sec:oiii_candidates}
To compute [O{\sc iii}] fluxes we used the previous results of calibration in Section~\ref{sec:Reduction}. We used a suited python script in determining the [O{\sc iii}] fluxes. With median flux value of all sources ($\widetilde{f}_{\rm all}$) and standard deviation ($\sigma_{\rm all}$), we took a pseudo-continuum from the simulated pseudo-spectra as part of pseudo-spectrum points by discarding ``high/low" values of the outliers: the sources with $f\,>\, \widetilde{f}_{\rm all}\,+\,2\,\times\,\sigma_{\rm all}$ (''high" outliers) or sources with $f\,<\,\widetilde{f}_{\rm all}\,-\,2\,\times\,\sigma_{\rm all}$ ("low" outliers). Hence, we defined the median as the continuum level ($f_{\rm cont}$) and its standard deviation ($\sigma_{\rm cont}$) to be the continuum noise (error). By finding the continuum subtracted flux (f$_{\rm cs}$([O{\sc iii}])) belonging to the peak in a range of $\pm\,2$ scan steps ($\pm\,20\,\AA$) within the predicted wavelength ($\lambda_{\rm pred}$) of the [O{\sc iii}] line and a TF effective transmission at the line wavelength (T([{O\sc iii}])), flux is computed as: 
\begin{equation}
 f([{\rm OIII}])=f_{\rm cs}([{\rm OIII}])\,\times\,T([{\rm OIII}]).
 \label{eq:flux_comp}
\end{equation}
Flux error is determined with the error propagation formula as:
\begin{equation}
\Delta\,f([{\rm OIII}])=T([{\rm OIII}])\,\times\,\sqrt{(\Delta\,f_{\rm b}([{\rm O III}]))^2\,+\,(\sigma_{\rm cont})^2},
 \label{eq:flux_error}
\end{equation}
where $\Delta\,f_{\rm b}([OIII])$ is a flux error before continuum subtraction.
We used Equation~\ref{eq:LO} to compute $\lambda_{\rm pred}$ using the redshift ($z$) derived with H$\alpha$ results and the rest frame wavelength ($\lambda_0=5007$ \AA):
\begin{equation}
 \lambda_{\rm pred}=\lambda_0\,\times\,(1+z).
 \label{eq:LO}
\end{equation}
The python script was applied to all the 152 preliminary [O{\sc iii}] candidates. While fluxes have been measured for 67/50 sources in centre/offset positions respectively with 23 duplicates. Finally, after visual inspection of pseudo-spectra we remained with 51 [O{\sc iii}] candidates.

\subsection{Derivation of H$\beta$ fluxes}
\label{sec:Hb_candidates}
In H$\beta$ case the TF collected flux is just the sum of the continuum (considerable absorption) and the H$\beta$ emission line. A total H$\beta$ emission is obtained by following a slightly similar process as for the [O{\sc iii}] case, but here involving the absorption correction. We first computed a continuum subtracted flux (f$_{cs}$(H$\beta$)) following the same procedure as for [O{\sc iii}].

\begin{figure}
	\centering
	\includegraphics[width=\columnwidth]{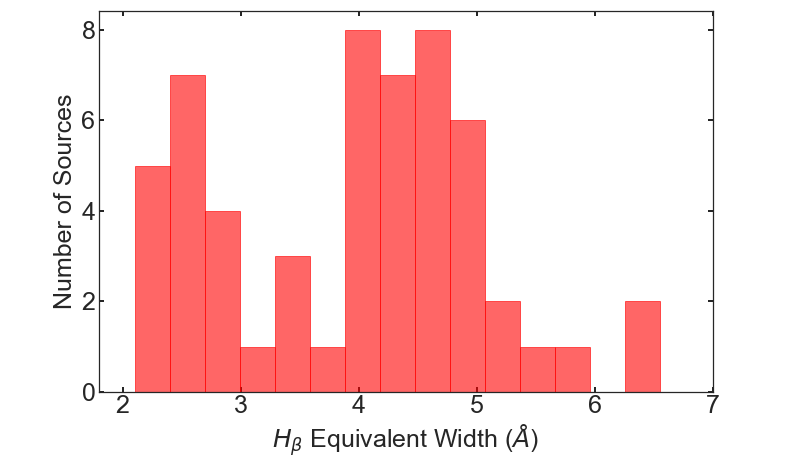}
	\caption{Distribution of H$\beta$ absorption line equivalent widths derived from the best-fitting templates and ages.}	
	\label{fig:absEW}		
\end{figure}

 For measuring the  equivalent width (EW) of H$\beta$, we went through the spectral energy distribution (SED) fitting, using the standard templates \citep[][hereafter BC03]{Bruzual2003} and {\tt LePhare} code \citep{Ilbert2006}. Since {\tt LePhare} makes use of the low-resolution composite stellar populations (CSP) while for measuring the EW(H$\beta$) we require a high-resolution SED templates, we obtained the best-fitting host galaxy stellar  template re-sampled at a finer resolution using the {\tt GALAXEV}\footnote{\tt http://www2.iap.fr/users/charlot/bc2003/} code (BC03). We derived the spectral evolution of the CSPs by integrating the equation of evolution of single stellar population (SSP) templates from BC03 with metallicities Z\,=\,0.0001, 0.0004, 0.004, 0.008 and 0.02 with star formation histories exponentially varying with time as SFR\,$\propto\,e^{-t/\tau}$, where $\tau$ ranges from 0.1 to 30.0 Gyr and taking the initial mass function (IMF) from \citet{Chabrier2003}. The output of this code includes the EW(H$\beta$) at different ages as defined in \citet{Trager1998}. Without applying the interpolation since both codes implement almost identical age steps, we have chosen the values of EW(H$\beta$) at the closest ages to those given by the best fit from {\tt LePhare}. The distribution of EW(H$\beta$) is given in Figure~\ref{fig:absEW} for the identified H$\beta$  lines, with the average ratio equivalent width being 4.0\,\AA.

The absorption line flux was measured by multiplying the value of EW(H$\beta$) by the previously derived pseudo-continuum. In average, the flux ratio is $f_{{\rm H_\beta,absorption}}/f_{{\rm H_\beta,emission}}\,\simeq\,0.43$. The next step is calculating the flux in absorption given by $f_{\rm abs}\,=\, {\rm EW_{H_\beta}}\,\times\,f_{\rm cont}$. The total flux (f$_{\rm total}$) is then finally computed as a sum of the continuum subtracted flux (f$_{\rm cs}$(H$\beta$)) and the flux in absorption as f$_{\rm total}(\rm H_\beta)$\,=\,f$_{\rm cs}$(H$\beta$)\,+\,f$_{\rm abs}$, and we determined the final line flux errors using the error propagation formula (now including absorption), including error in absorption as: 
\begin{equation}
\Delta\,f_{\rm total}(\rm H\beta)=\sqrt{(\Delta\,f(\rm H\beta))^2+(\sigma_{\rm cont}\,\times\,EW(\rm H\beta))^2}
 \label{eq:flux_error2}
\end{equation}
We obtained flux measurements for 69 and 75 sources in central and offset positions, respectively, with 28 duplicates. After the visual inspection of pseudo-spectra, we identified 80 final candidates for H$\beta$ emission. 

\subsection{Analysis of individual objects and selection of the final sample of [O{\sc iii}] and H$\beta$ emitters}

\begin{figure*}
 \centering
 \includegraphics[width=18cm]{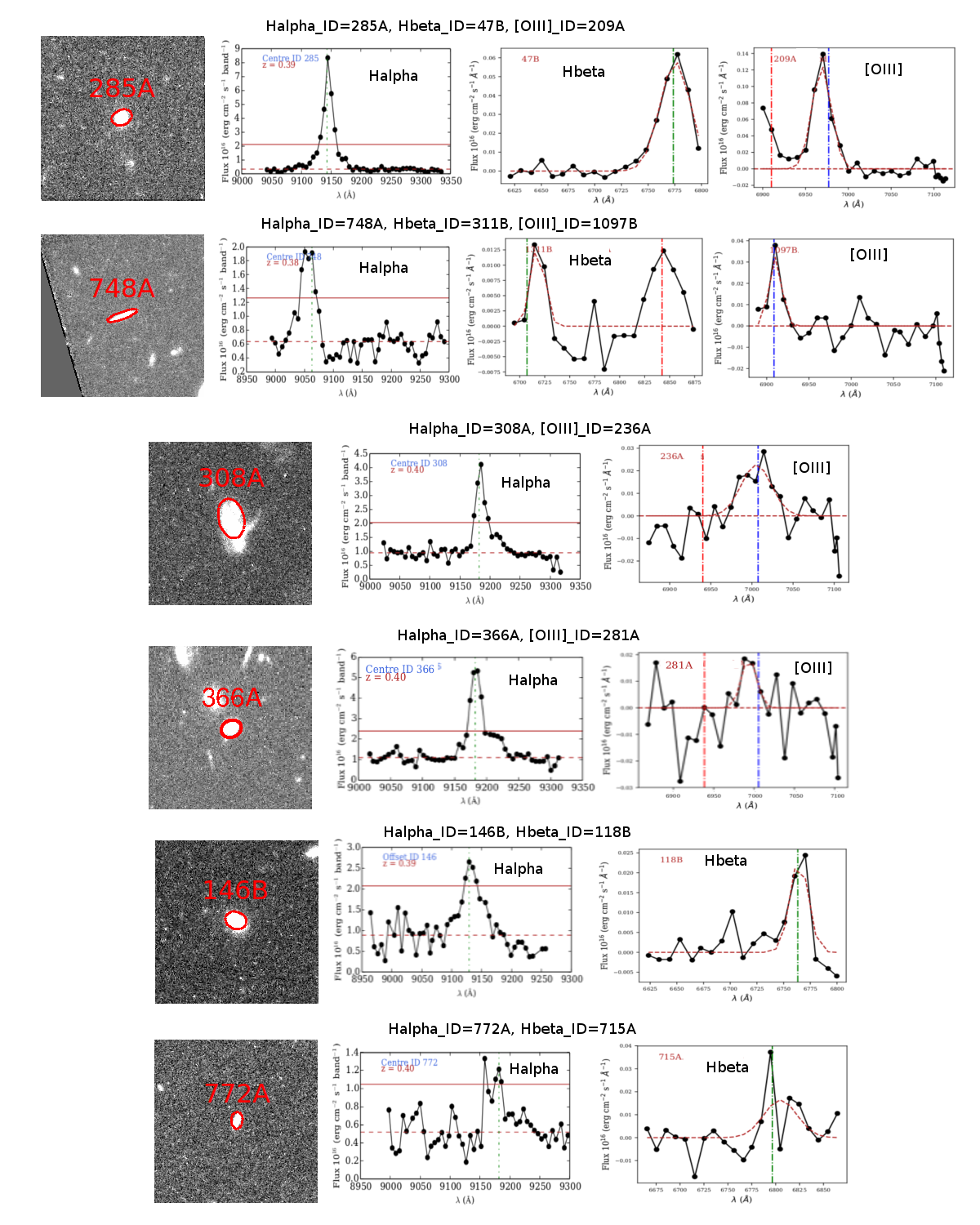}
    \caption{Example of HST/ACS thumbnails (left) and pseudo-spectra of ELGs with H$\alpha$ (second column), and [O{\sc iii}] and/or H$\beta$ emissions (third and/or fourth columns, as indicated on each pseudo-spectrum). The predicted H$\alpha$ position with spectroscopic redshifts \citep{Moran2007} and the predicted H$\beta$ line wavelength positions derived with H$\alpha$ redshifts are represented with green vertical lines in their corresponding pseudo-spectra, while the predicted wavelength positions for [O{\sc iii}]5007 and [O{\sc iii}]4959 both computed with H$\alpha$ are indicated with blue and red vertical lines, respectively, when detected.}
    \label{fig:EL_sample}
  \end{figure*}

 We obtained in total 51 [O{\sc iii}] and 80 H$\beta$  emission line final candidates with 29 common candidates for both lines resulting with a total of 102 unique candidates, as described in Sections~\ref{sec:oiii_candidates} and~\ref{sec:Hb_candidates}. For final sample selection and to make sure that fluxes are coming from  single sources, we first went through the visual inspection of thumbnails from HST data described in Section~\ref{sec:ancillary_data}. In total, we found thumbnails of 31 and 52  [O{\sc iii}] and H$\beta$  candidates, respectively, with 19 sources in common. Most of the sources ($\sim$\,91\%) were inspected to be well isolated. Secondly, for remaining sources we inspected the H$\alpha$  pseudo-spectra \citep{Sanchez2015}, and removed those sources with line offset of $\geq 10$ \AA\ and/or those sources with no clearly detected peak. We finally detected all targeted emission lines ([O{\sc iii}], H$\beta$, [N{\sc ii}] and H$\alpha$) for 20 ELGs. Our final sample is summarized in Table~\ref{tab:final_lines}.
\begin{table}
	\centering
	\caption{The number of preliminary candidates for [O{\sc iii}] and/or H$\beta$ emission; with final detected emission line sources.}
	\label{tab:final_lines}
	\begin{tabular}{lcccc} 
		\hline
	        \hline
Sources & [O\sc iii] & H$\beta$ & [O{\sc iii}] and & [O{\sc iii}]\,+\,H$\beta$\,+\, \\
	      & 		&		& H$\beta$ & [N{\sc ii}] + H$\alpha$ \\
		\hline
Pre-candidates & 51 & 80 & 29 & 28 \\
Final emitters & 35 & 59 & 21 & 20 \\
	\hline
	\hline
\end{tabular}
\end{table}
Figure~\ref{fig:EL_sample} gives an example of sources with [O{\sc iii}] and/or H$\beta$ lines, showing the ACS thumbnails and available pseudo-spectra. 
	
\section{Analysis and Results}
\label{sec:Results}
\subsection{Redshift, flux and luminosity distributions of [O{\sc iii}] and H$\beta$ line emitters}
\begin{figure}
\centering
    \includegraphics[width=\columnwidth]{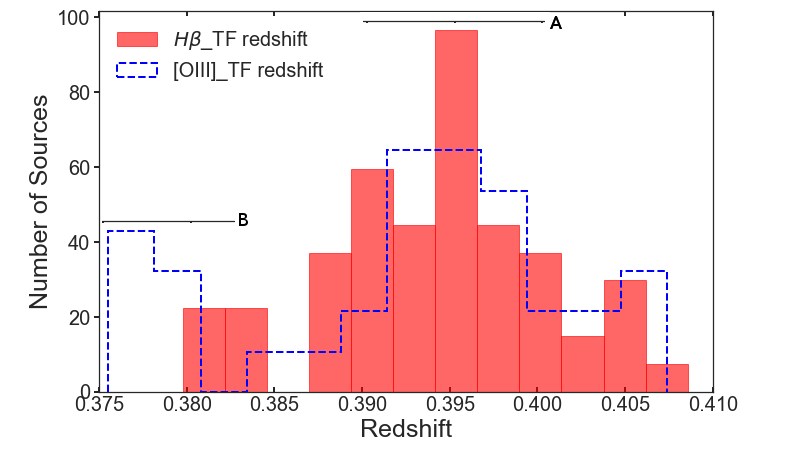}
    \includegraphics[width=\columnwidth]{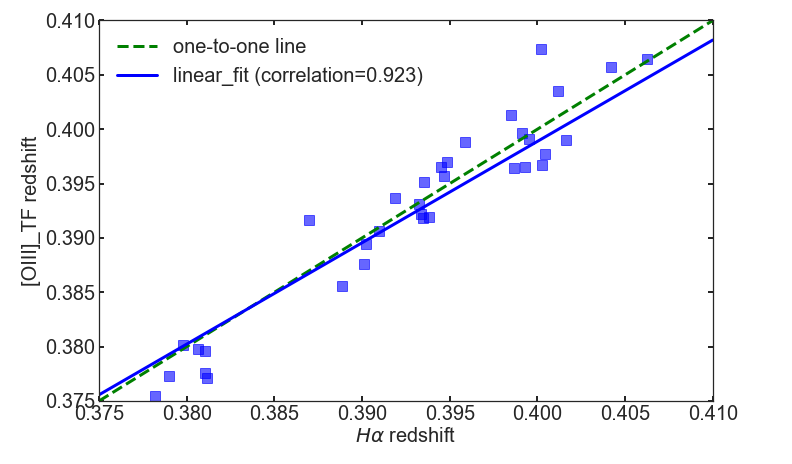}
    \includegraphics[width=\columnwidth]{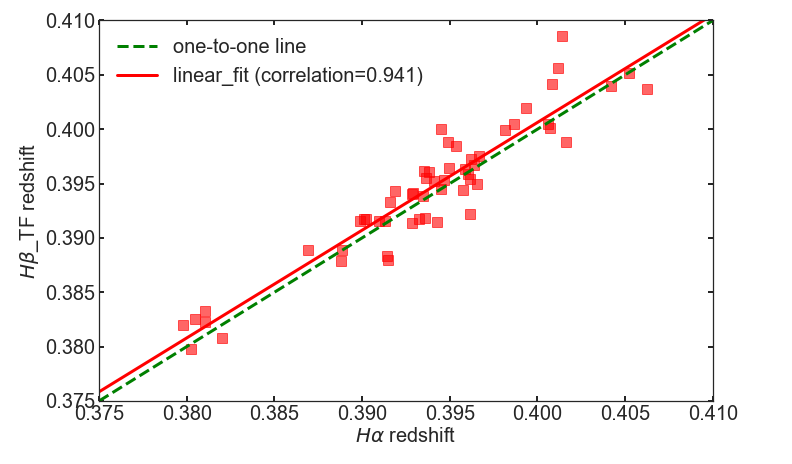}
        \caption{Redshift distribution of [O{\sc iii}] and H$\beta$ ELGs (top plot). Two dynamical structures A and B can be seen, as suggested in previous works \citep{Czoske2002,Sanchez2015}. In addition, linear correlation between [O{\sc iii}] (middle plot) and H$\beta$ (bottom plot) TF derived redshifts and H$\alpha$ redshifts has been represented.}
    \label{fig:redshift1}
   \end{figure}

\begin{figure}
\centering
    \includegraphics[width=\columnwidth]{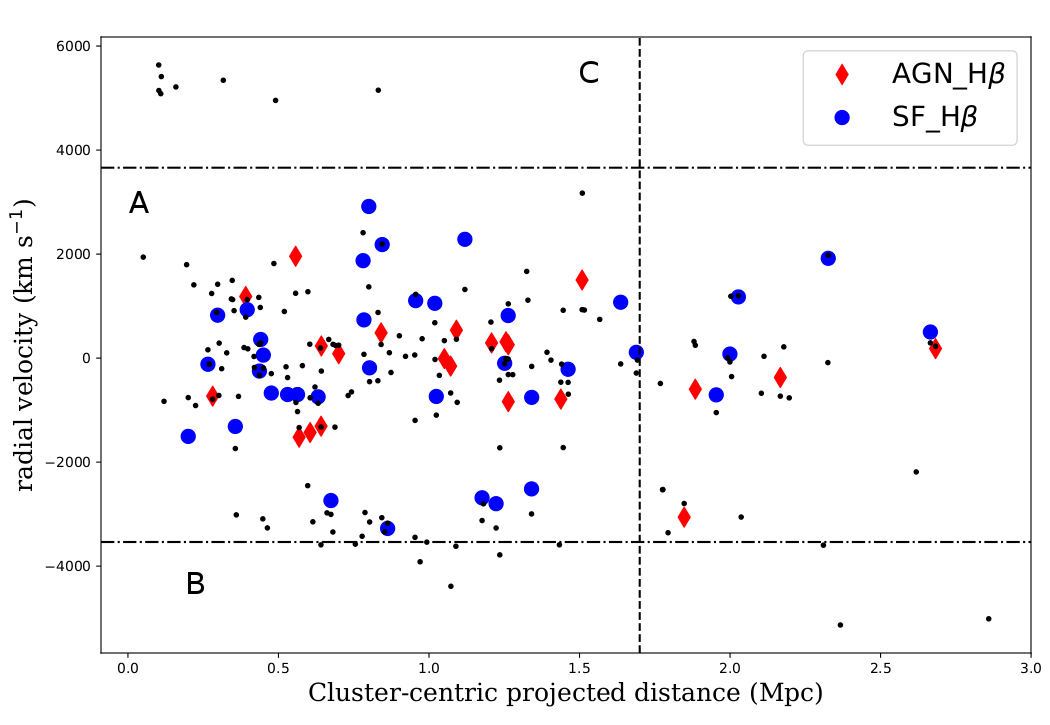}
    \includegraphics[width=\columnwidth]{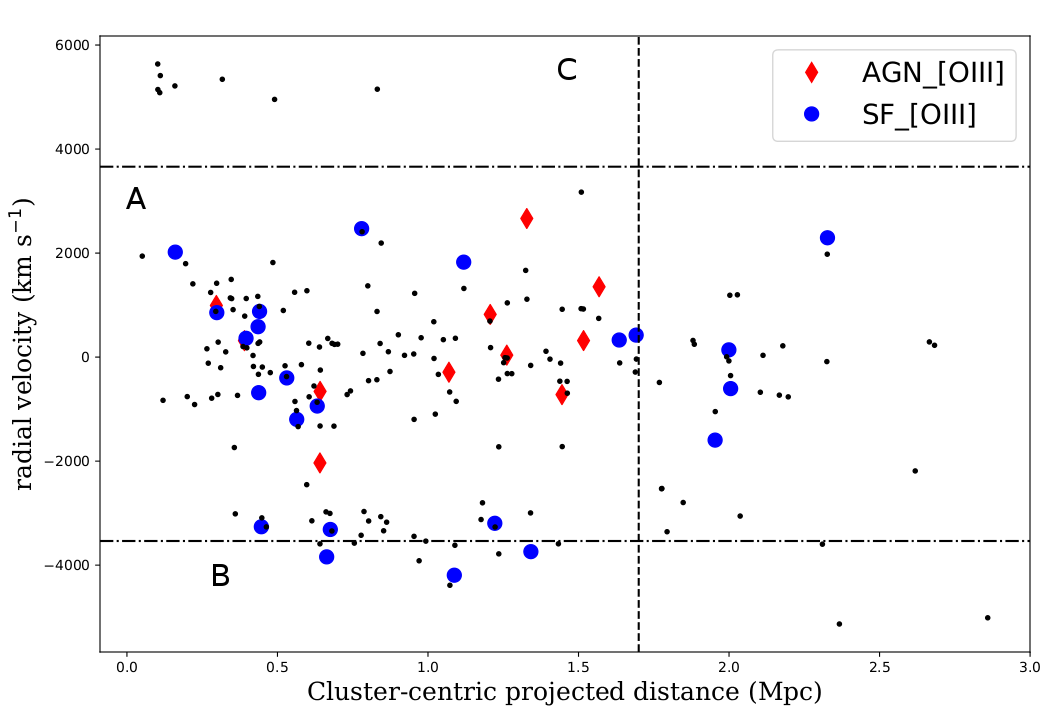}
       \caption{Radial velocity versus clustercentric distance for H$\beta$ (top plot) and [O{\sc iii}] (bottom plot) ELGs. Red diamonds represent AGNs while blue dots correspond to SF galaxies. The virial radius (r$_{\rm vir}=1.7${\rm Mpc}) is represented by dotted vertical line \citep{Treu2003}. The dashed-dotted horizontal lines stand for the radial velocity limits fully covered within the field of view of the two OSIRIS TF pointings. The small black points correspond to ELGs from \citet{Sanchez2015} with H$\alpha$ emission. The two reported structures of the cluster are  marked with ``A'' and ``B'', whereas the putative possible component is represented with letter ``C''.}
    \label{fig:velocity}
   \end{figure}

Redshifts of nearly spectroscopic-quality were derived  in \cite{Sanchez2015} from the positions of the H$\alpha$ emission lines for all our ELGs. We computed the redshifts through detected [O{\sc iii}] and H$\beta$ lines and then compared our results with H$\alpha$ redshifts presented in \citet{Sanchez2015}. In Figure~\ref{fig:redshift1} we show the distribution of redshifts as compared with H$\alpha$ measurements. The relative redshift deviations, defined as $\lvert\,{\rm z_{TF}}\,-\,{\rm z_{H\alpha}}\rvert\,/\,(1 + {\rm z_{H\alpha}})$, are  $\sim$\,0.002 and $\sim$\,0.001 for [O{\sc iii}] and H$\beta$, respectively, confirming therefore the spectroscopic quality of the TF-derived redshifts. Our redshift distributions are in line with previous results \citep{Czoske2002,Sanchez2015}, with average redshifts of $\sim$\,0.392 and $\sim$\,0.394 for [O{\sc iii}] and H$\beta$  lines, respectively. Moreover, we recover two dynamical structures represented by double-peaks in the redshift distribution, where structure ``A'' (centred at z $\simeq$ 0.395) corresponds to the main cluster, while structure ``B'' (centred at z $\simeq$ 0.380) represents the infalling component \citep{Czoske2002,Kneib2003,Moran2007,Sanchez2015}. Our redshift measurements are also in good agreement with those obtained from H$\alpha$ data \citep{Sanchez2015}, as can be seen in Figure~\ref{fig:redshift1} (middle and bottom plots), with linear correlation factors of 0.923 and 0.941, respectively. As determined by \citet{Czoske2002}, the velocity distribution of structure A was determined to be regular with a dispersion of $\sim$\,600\,\textrm{km}\,\textrm{s}$^{-1}$ at projected distance $>3'$, while the formal velocity dispersion within $5'$ is $\sigma_{cent}$\,=\,1050\,\textrm{km}\,\textrm{s}$^{-1}$ for 193 galaxies. The environment conditions were parametrised by the local surface density computed by P\'erez-Mart\'inez et al. (in prep.) from the distance to the 5th-neighbour ($\Sigma_5$), according to \cite{Dre85}. Moreover, we computed the radial velocity of each member galaxy and derived the projected clustercentric distance separately for H$\beta$ and [O{\sc iii}] emitters to produce and discuss on the projected phase--space diagrams shown in Figure~\ref{fig:velocity}.
   
\begin{table}
	\centering
	\caption{The statistical distribution of fluxes of [O{\sc iii}] and H$\beta$ emitters in $\mathrm{erg}\,\mathrm{s}^{-1}\,\mathrm{cm}^{-2}$; with Q$_1$ being first quarter, Q$_3$ being third quarter and Med representing median.}
	\label{tab:flux}
	\addtolength{\tabcolsep}{-2.5pt}
	\begin{tabular}{ccccc} 
		\hline
	        \hline
Emission  & Q$_1$  & Med  & Q$_3$ & Adopted  \\
line & & & & flux limit \\ 
		\hline
[O\sc iii] & $3.7\times10^{-17}$ & $6.2\times10^{-17}$ & $1.0\times10^{-16}$ & $2.5\times10^{-17}$ \\
H$\beta$   & $4.1\times10^{-17}$ & $1.3\times10^{-16}$ & $7.4\times10^{-17}$ & $2.0\times10^{-17}$ \\
	\hline
	\hline
\end{tabular}
\end{table}

\begin{table}
	\centering
	\caption{The statistical distribution of logarithm luminosities (Q$_1$, Med, Q$_3$) in $\mathrm{erg}\,\mathrm{s}^{-1}$ of [O{\sc iii}] and H$\beta$ emitters.}
	\label{tab:lumin}
	\begin{tabular}{lccc} 
		\hline
	        \hline
Emission line & Q$_1$  & Med  & Q$_3$  \\
		\hline
[O\sc iii] & 40.3 & 40.5 & 40.7 \\
H$\beta$   & 40.3 & 40.6 & 40.9 \\
	\hline
	\hline
\end{tabular}
\end{table}

\begin{figure}
\centering
    \includegraphics[width=\columnwidth]{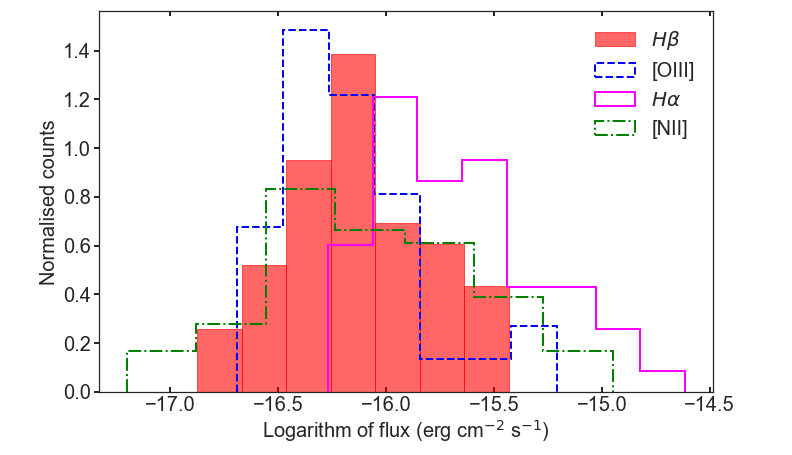}
    \includegraphics[width=\columnwidth]{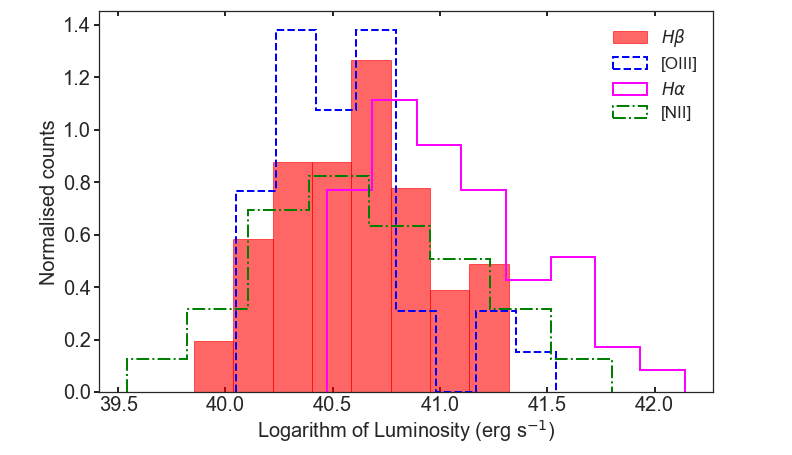}
    \caption{Comparison of the flux (top plot) and luminosity (bottom plot) of H$\beta$ (red filled histogram), [O{\sc iii}] (blue dashed line), H$\alpha$ (violet solid line), and [N{\sc ii}] (green dash-dotted line) ELGs.} 
    \label{fig:flux_lumin_compare}
\end{figure}

The statistics of the distributions of  fluxes and luminosities of [O{\sc iii}] and H$\beta$ are given in Tables~\ref{tab:flux} and~\ref{tab:lumin}. A comparison of these distributions with those of H$\alpha$ and [N{\sc ii}] is depicted in Figure~\ref{fig:flux_lumin_compare}. We found that [O{\sc iii}] and H$\beta$ fluxes and luminosities are on average comparable with H$\alpha$ and [N{\sc ii}] values, although slightly lower. As a further comparison, median fluxes of H$\alpha$ and [N{\sc ii}] are $1.9\times10^{-16}\,\mathrm{erg}\,\mathrm{s}^{-1}\,\mathrm{cm}^{-2}$ and $7.7\times10^{-17}\,\mathrm {erg}\,\mathrm {s}^{-1}\,\mathrm{cm}^{-2}$, respectively. Median logarithmic luminosities ($\log\,\mathrm{L}$) are $41.0\,\mathrm{erg}\,\mathrm{s}^{-1}$ and $40.6\,\mathrm{erg}\,\mathrm{s}^{-1}$, respectively. 

\subsection{Morphologies of the [O{\sc iii}] and H$\beta$  emitters}
\label{sect:lines_morpho}

In this work we used morphological classification from \citet{Amado2019}. Moreover, \citet{Treu2003} and \citet{Moran2007} have previously carried out a visual classification within 5\,Mpc of the cluster centre. 
In all these works galaxies have been classified between early-type (ET) and late-type (LT). After cross-matching our data with all the three references, we found morphological classes for 24 [O{\sc iii}] emitters (8 ET and 16 LT), and 41 H$\beta$ sources (13 ET and 28 LT), leaving significant number of sources without classification. This could be because \citet{Amado2019} classification is only within 1\,\textrm{Mpc} and others give classification only for brightest sources. This gives us similar fractions of ETs (33\% for [O{\sc iii}] and 32\% in case of H$\beta$), and LTs (67\% for [O{\sc iii}] and 68\% for H$\beta$). 

As expected, we found a larger populations of LT galaxies, and on average their [O{\sc iii}] and H$\beta$ emission lines are similar to those of ETs. Median logarithmic luminosities (log L) of LT and ET are 40.59 \,\textrm{erg}\,\textrm{s}$^{-1}$ and 40.54 \,\textrm{erg}\,\textrm{s}$^{-1}$ in the case of [O{\sc iii}], and 40.68\,\textrm{erg}\,\textrm{s}$^{-1}$ and 40.62 \,\textrm{erg}\,\textrm{s}$^{-1}$ in the case of H$\beta$, respectively.

\subsection{AGN versus SF classes: BPT diagrams}
\label{sect:BPT_analysis}
\begin {figure}
\centering
  \includegraphics[width=\columnwidth]{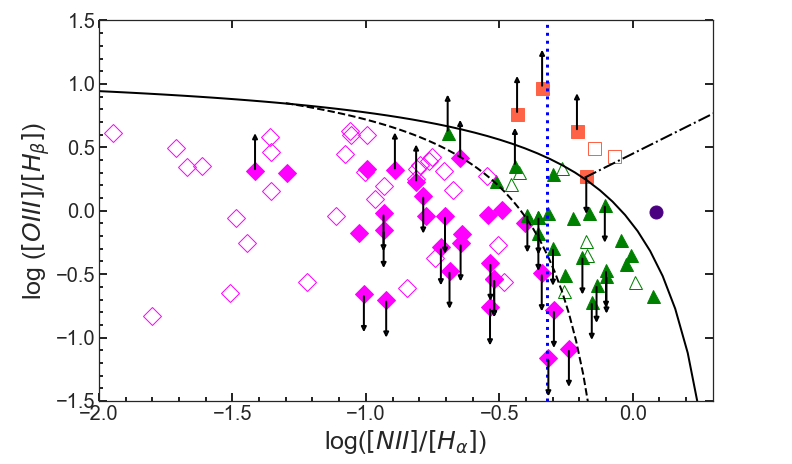}
    \caption{BPT diagram for separating SF (magenta), composite (green), Seyfert--2 (red) and AGN--LINER (indigo) ELGs. Filled symbols represent the cluster sources and the open ones the field galaxies. The arrows correspond to the cluster ELGs with either [O{\sc iii}]- or H$\beta$-only line detection (upward for [O{\sc iii}], while downward for H$\beta$). Continuous, dashed, and dot-dashed lines correspond to the boundaries given by \citet{Kewley2006}, \citet{Kauffmann2003} and \citet{Schawinski2007}, respectively. The vertical dotted line marks the SF-AGN separation, as prescribed by \citet{Kauffmann2003}. More details are given in text.}
    \label{fig:BPT}
\end{figure}

 For the sources with all the four lines detected (as described in Table~\ref{tab:final_lines}), we checked their spectral type using the BPT-NII diagnostic diagram \citep{Baldwin1981,Kewley2006}, separating galaxies into star-forming (SF), composite and AGN (Seyfert--2 or LINERs), as shown in Figure~\ref{fig:BPT}. As a previous step, we segregated the 15 sources in our sample classified as broad-line AGNs (BLAGN) in \citet{Sanchez2015}.  We identified 9 ELGs with [O{\sc iii}], [N{\sc ii}] and H$\alpha$ detections and H$\beta$ non detected, as well as 28 ELGs with H$\beta$, [N{\sc ii}] and H$\alpha$ flux measurements but without detection of the [O{\sc iii}] line. In order to include these sources in the BPT-NII plot we used the flux limits shown in Table \ref{tab:flux} to assign upper bounds to non-detections, which are represented with upward or downward arrows in this figure. The boundary between SF and composite galaxies was taken from \citet{Kauffmann2003}, and between composite and AGNs from \citet{Kewley2001}. To separate between LINERs and Seyfert--2 galaxies we used the prescription from \citet{Schawinski2007}. The minimum and maximum limits (for [O{\sc iii}]/H$\beta$ and [N{\sc ii}]/H$\alpha$) were adapted from \citet{Kewley2006}. Although all of these boundaries have been established using a sample of galaxies with redshifts z\,<\,0.1, and although our cluster galaxies are at z\,$\sim$\,0.4, we consider no significant evolution taking into account results of \citet{Kewley2013} verifying that the relation remains valid up to z\,$\sim$\,1. 

In our cluster sample, we found a total of 29 (51\%) SF galaxies (8 with all emission lines, 4 with no H$\beta$ detection, 17 with no [O{\sc iii}] detection), 23 (40\%) composites (11 with all emission lines, 2 with no H$\beta$ detection, 10 with no [O{\sc iii}] detection), 1 (2\%) LINER galaxy with all emission lines and 4 (7\%) Seyfert--2 ELGs (1 with no [O{\sc iii}] detection and 3 with no H$\beta$ detection). The results of our classification are summarised in Table~\ref{tab:class}. 

\begin{table}
	\centering
	\caption{Summary of our spectral classification of galaxies in ZwCl0024+1652 cluster using BPT diagram as shown in Figure~\ref{fig:BPT}.}
	\label{tab:class}
	\addtolength{\tabcolsep}{-2.5pt}
	\begin{tabular}{lcccc} 
		\hline
	        \hline
\textbf{Emission}  & \textbf{SF}  & \textbf{Comp.} & \textbf{AGN(Sy2+LINER)} & \textbf{Total}  \\
		\hline
No H$\beta$ detection & 4 & 2 & 3 & 9 \\
No [O{\sc iii}] detection  & 17 & 10 & 1 & 28 \\
All lines detected & 8 & 11 & 1 & 20 \\
Total & 29 & 23 & 5 & 57 \\
	\hline
	\hline
\end{tabular}
\end{table}
Additionally, we selected all the galaxies with detected emission in the chemical species cited above from the VVDS-Deep field sample described in Section~\ref{sec:field_data}. From 177 field sources we identified 41 sources having all the four lines. We segregated 32 (78\%) SF galaxies, 7 (17\%) composite, and 2 (5\%) Seyfert--2 galaxies, also represented in Figure~\ref{fig:BPT} (open symbols). According to this, with the caveat regarding the small-number statistics, there are $\sim$\,1.5 times more SF galaxies in the field and almost one-third less AGN+composite galaxies than in cluster galaxies. These differences are about three times the mean uncertainty derived from a Poissonian statistics derived from the absolute numbers of field galaxies belonging to each spectral type. 

\begin{figure*}
	\centering
	\includegraphics[width=19cm]{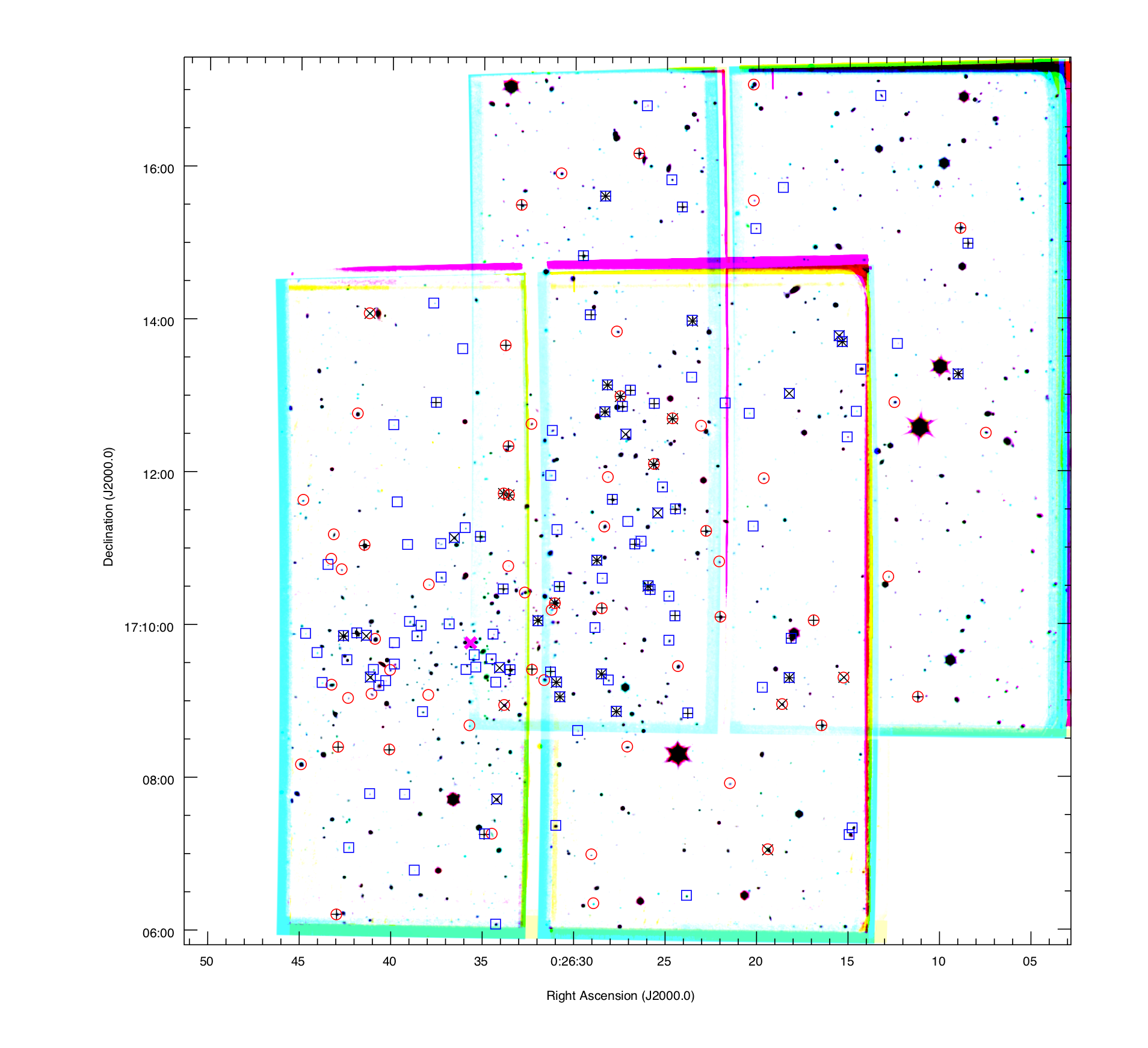}
	\caption{RGB mosaic (H$\alpha$, [O{\sc iii}], and H$\beta$, respectively) of the OSIRIS/GTC deep images of ZwCl0024+1652. Blue squares correspond to SF galaxies and red circles represent AGN hosts, as provided by \citet{Sanchez2015}. The ``X'' symbols denote [O{\sc iii}] emitters while ``+'' signs represent H$\beta$ ELGs from our work. The magenta cross (\textcolor{magenta}{$\times$}) denotes the centre of the cluster (galaxies/BCG).}
	\label{fig:spatial_distribution}		
\end{figure*}

Hereafter, and in order to overcome the limitation imposed by the need of simultaneous detection of both the [O{\sc iii}] and H$\beta$ lines, we will use the simplified criterion applied by \cite{Sanchez2015} to separate SF galaxies from AGN hosts, based on the prescription of \citet{Kauffmann2003},  namely the boundary defined by the flux ratio [N{\sc ii}]/H$\alpha$\,$=$\,0.478. With this assumption, the analysis and discussion described in the following sections is based on a segregated total of 35 SF galaxies and 22 AGN hosts. We compared our results here with statistics from \citet{Sanchez2015}. Out of 174 ELGs in \citet{Sanchez2015}, 60 (34\%) were AGN hosts while 114 (66\%) were found to be SF galaxies. Our work gave similar results with nearly comparable distribution: 38.5\% AGN hosts with 61.5\% SF galaxies. The sky distribution of H$\alpha$ sample including our [O{\sc iii}] and H$\beta$ results is presented in Figure~\ref{fig:spatial_distribution}. The sky positions of the ELGs is described on a blended image of an inverted RGB composed of deep images determined as in Section~\ref{sec:Reduction} with H$\alpha$ image. Finally, applying the same rule to the VVDS-Deep population, we segregated a total of 39 SF field galaxies and 7 AGN hosts. A broader discussion on comparison between cluster and field galaxies is given in Section~\ref{sec:comparison_FC}. 

\subsection{Extinction-corrected star formation rate}
\label{sect:extinction_corrected_SFR}
\begin{figure}
	\centering
	\includegraphics[width=\columnwidth]{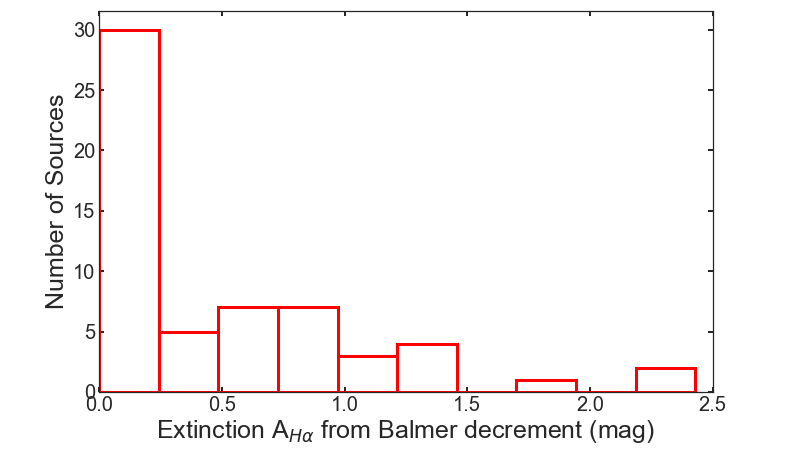}
	\caption{Distribution of extinctions derived from the Balmer decrement, assuming a galactic extinction law of \citet{Seaton79}.} 
	\label{fig:extinction_Seaton}		
\end{figure}

We used the total H$\beta$  flux, measured as the sum of the emission and absorption components computed as explained earlier, to compute the extinction correction from the Balmer decrement $f_{H\alpha}/f_{H\beta}$ using Equation~\ref{eq:extinction} below: 
\begin{equation}
\label{eq:extinction}
A_{\mathrm{H}\alpha} = \frac{2.5}{k_{\rm H\alpha}/k_{\rm H\beta}-1}\log\left(\frac{1}{2.85}\frac{f_{\rm H\alpha}}{f_{\rm H\beta}}\right)
\end{equation}

\noindent where $k_{\rm H\alpha}/k_{\rm H\beta}$\,=\,1.48 for the galactic extinction law from 
\cite{Seaton79}.  
The distribution of the computed extinction values is represented in Figure~\ref{fig:extinction_Seaton}. The largest peak corresponds to unabsorbed galaxies (A$_{\rm H\alpha}$\,$\simeq$\,0). This could be expected since, due to the incompleteness of the H$\beta$ sample (the H$\beta$ line is actually detected for 59 galaxies ($\sim$\,34\%) out of 174 H$\alpha$ emitters), our observations favour the detection of strong H$\beta$ emitters, i.e. intrinsically luminous sources or unabsorbed ones. On the other hand, a second peak close to A$_{\rm H\alpha}\simeq 0.75$ is observed, which is consistent with the usual assumption ($\sim$\,1 mag) after \citet{Kennicutt1992}.

\begin{figure}
	\centering
	\includegraphics[width=\columnwidth]{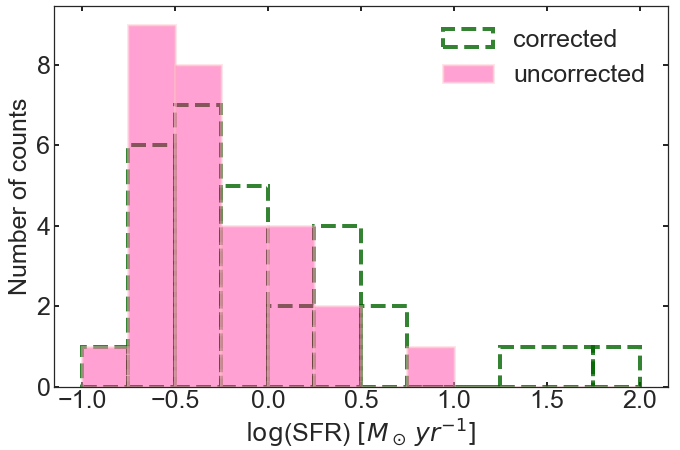}
	\includegraphics[width=\columnwidth]{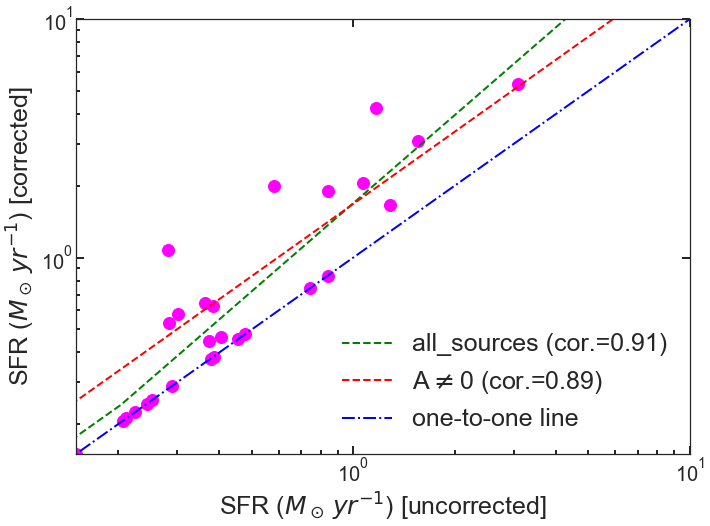}
	\caption{\textbf{Top plot:} Distribution of corrected and uncorrected SFR. \textbf{Bottom plot:} Relation between extinction corrected and uncorrected SFR for ZwCl0024+1652, according to the details given in Section~\ref{sect:extinction_corrected_SFR}. Blue dash-dotted and green dashed lines show one-to-one correlation and linear fit for all sources, respectively. While the red dashed line represents the linear fit for those galaxies with zero extinction.} 
	\label{fig:SFR_relations}		
\end{figure}

The average uncertainty of the Balmer decrement obtained by error propagation from Equation~\ref{eq:extinction} is about 0.88 mag. Its relative error is 27.1/37.5\% (median/mean). Once we derive the extinction correction, the SFR can be computed using a suitable scaling relation, as given by \cite{Kennicutt1998}:

\begin{equation}
\label{eq:sfr_kennicutt}
{\rm SFR} (M_{\odot}yr^{-1}) = \frac{10^{\rm A_{H\alpha}/2.5}}{1.58}  7.94 \times 10^{-42} L(\rm H\alpha) ({\rm erg\ s^{-1}}),
\end{equation}

\noindent where the factor 1.58 corresponds to the conversion from Salpeter to \cite{Chabrier2003} initial mass function (IMF). The distributions of uncorrected  and corrected H$\alpha$ SFR are depicted in the top panel of Figure~\ref{fig:SFR_relations}.

The errors in both the uncorrected and corrected SFR have been computed as usual by propagation. The average uncertainty in the uncorrected SFR is 19.5/20.9\% (median/mean). However, given the large uncertainty in A$_{\rm H\alpha}$, the fractional errors in the corrected SFR are much larger, 67.3/71.2\% (median/mean) in our case.

The correlation between the uncorrected and corrected SFR is depicted in the bottom panel of Figure~\ref{fig:SFR_relations}. Even though the results should be taken with caution, given the large uncertainties involved, a quite clear correlation is observed. Moreover, if  the range is restricted to SFR$_{\rm H\alpha,corrected}$\,$<$\,12\,M$_{\odot}$ yr$^{-1}$, thus avoiding two objects that are luminous and highly absorbed (A$_{\rm H\alpha}$\,=\,2.43\,mag), a reasonable linear correlation equal to 0.91 with slope of 1.23 is observed considering all the sources including sources with A$_{\rm H\alpha}$\,=\,0 (green dashed line in Figure~\ref{fig:SFR_relations}, bottom plot); while excluding the sources with A$_{\rm H\alpha}$\,=\,0, the correlation factor is changed slightly to 0.89 with 1.00 being a new slope (red dashed line in Figure~\ref{fig:SFR_relations}, bottom plot). We performed a quantitative comparison between the two (uncorrected and corrected SFR) by using Kolmogorov--Smirnov two sample (KS2) test. We computed the p-value of our data to be 0.28, showing that both are coming from the same distribution. 

\section{Discussion}
\label{sec:Discussions}
\subsection{Morphological and spectral classes of the cluster members with [O{\sc iii}] and H$\beta$ emission}
\label{sec:morpho_bpt}
\begin{figure}
\centering
    \includegraphics[width=\columnwidth]{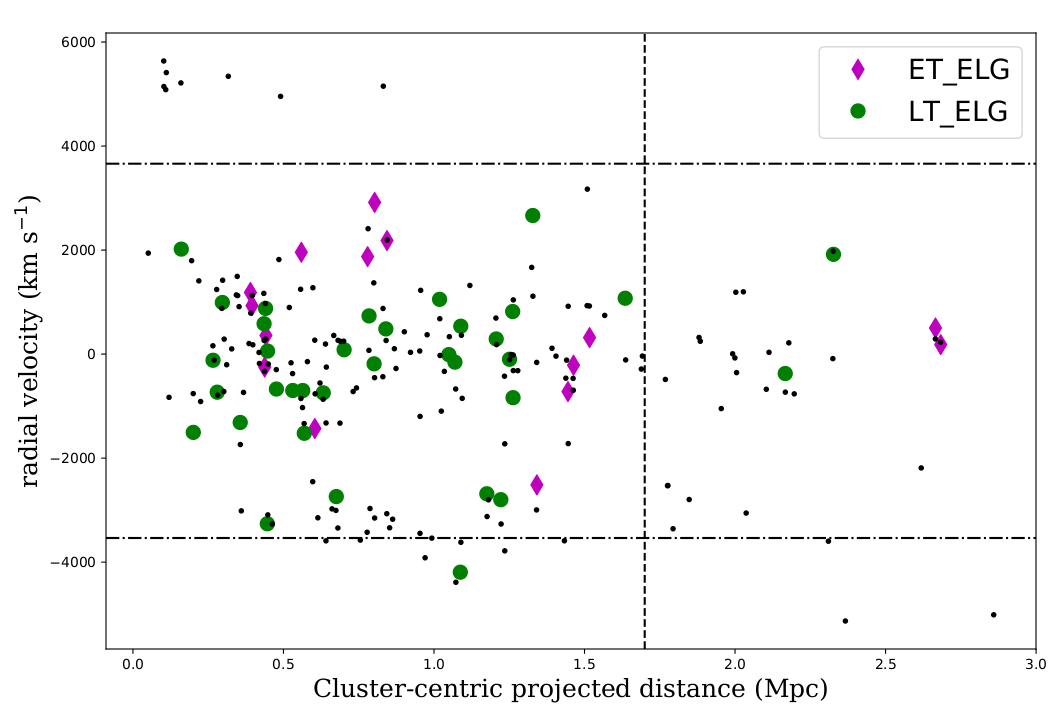}
    \caption{Radial velocity versus clustercentric distance for H$\beta$ and [O{\sc iii}] ELGs with morphological classifications. Magenta diamonds represent ETs while green dots correspond to LT galaxies. The virial radius (r$_{\rm vir}=1.7${\rm Mpc}) \citep{Treu2003} is represented by dotted vertical line. The dashed-dotted horizontal lines stand for the radial velocity limits fully covered within the field of view of the two OSIRIS TF pointings. The small black points correspond to ELGs with H$\alpha$ emission from \citet{Sanchez2015}.}
    \label{fig:velocity_morpho}
   \end{figure}

\begin{figure}
	\centering
	\includegraphics[width=\columnwidth]{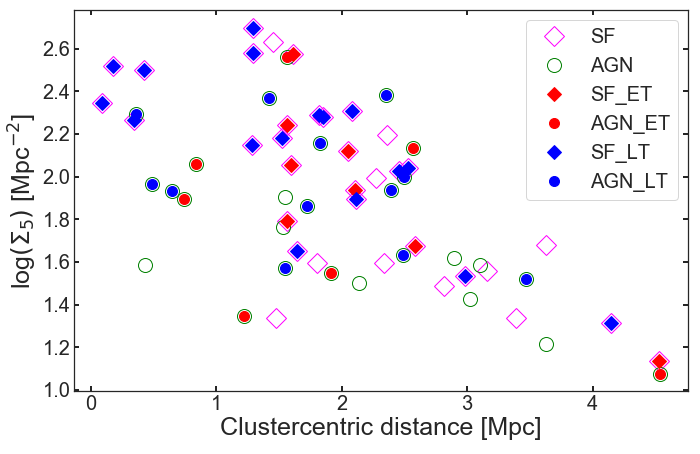}
	\caption{Dependence of the local density on the clustercentric distance of all SF galaxies (magenta open diamonds) and AGN (green open circles) that belong to the main cluster structure. Blue and red symbols stand for LT and ET galaxies, respectively.}
	\label{fig:sigma_r}
\end{figure}

In \citet{Amado2019} it was discussed in detail that the morphological fractions of ZwCl0024+1652 member galaxies vary with clustercentric distance in such a way that ET galaxies dominate over the LTs at least up to a clustercentric distance of $\sim$\,1\,Mpc. 
Other studies also suggested that the cores of nearby clusters have lower LT than the ET fraction, and that LT fraction increases with clustercentric distance while the ET fraction decreases \citep[e.g.,][]{Whitmore1993,Fasano2012,Fasano2015,PC2016,Kelkar2019}. 
In addition to this, \cite{Oh2018} showed that beside ET galaxies, passive spirals can also be found in highly dense regions, such as cluster centres, suggesting that they have gone through environmental quenching. 
Our TF [O{\sc iii}] and H$\beta$ observations result with detection of the brightest sources. 
As showed in Section~\ref{sect:lines_morpho}, despite small sample our results show that LT galaxies dominate over the ETs when dealing with ELGs, as expected. 
In addition to this, our results are in general in line with previous ones where LT galaxies show to be in average at larger clustercentric distances \citep[e.g.,][]{nantais2013,Fasano2015}. 
We found that median clustercentric distances of LT and ET ELGs are 1.82\,Mpc and 1.60\,Mpc when dealing with [O{\sc iii}] emitters and 1.93\,Mpc and 1.61\,Mpc in the case of H$\beta$ emitters, respectively. 
We also present a phase--space relation of H$\beta$ and [O{\sc iii}] ELGs separated with morphology in Figure~\ref{fig:velocity_morpho}. Apart from small statistics, from Figure~\ref{fig:velocity_morpho} we deduce that the inner most region of the cluster (projected clustercentric distance $\lesssim$\,0.3\,Mpc) is almost devoid of ELGs with morphologies (ET as well as LT).
Moreover, from Figure~\ref{fig:sigma_r} taking the morphological distributions in relation to radial distance, we can be see that LT ELGs are more dispersed along the clustercentric distance than ETs. 
In addition to the results obtained in \citet{Amado2019} we can now see from the current work using TF [O{\sc iii}] and H$\beta$ data that ET galaxies with signs of either AGN or SF activity seem to avoid the most central parts of ZwCl0024+1652 cluster. 

Another point that we raise here is the distribution of spectral types with in the cluster. Previous works showed that cores of regular galaxy clusters are characterised by high density while density decreases outwards with clustercentric distance \citep[e.g.,][]{Treu2003,Weinmann2006,Fasano2015}. Our results presented in Figure~\ref{fig:sigma_r} agree with these results. When representing all ELGs that belong to the principal cluster structure, anti-correlation can be observed between $\Sigma_5$ and clustercentric distance. This monotonic relation of local density to the clustercentric distance indicates that our cluster is a regular one. Taking into account Figure~\ref{fig:sigma_r}, and a small number of SF and AGN sources in the most central part of cluster, our results are in line with previous findings which suggest that the number of SF galaxies \citep[e.g.,][]{Gomez2003,Christlein2005,Hansen2009,Linden2010} and AGN \citep[e.g.,][]{Gavazzi2011,Davies2017,Marshall2018,Li2019,Koulouridis2019} decreases towards the cluster centre. The majority of SF galaxies and AGN found in this work are distributed at clustercentric distances $>$\,1.3\,Mpc. Moreover, the results obtained by \citet{Deshev2017} while analysing the evolution of galaxies in merging clusters with special focus on A520 at z\,=\,0.2, indicate that the core of such clusters up to a clustercentric distance of 1.5\,Mpc is almost devoid of SF galaxies. Since it is well known that ZwCl0024+1652 cluster is a merging cluster \citep{Czoske2002,Jee2007}, our findings confirm the existing results.   

\subsection{Colour and stellar masses}
\label{sec:mass}
\begin{figure}
	\centering
	\includegraphics[width=\columnwidth]{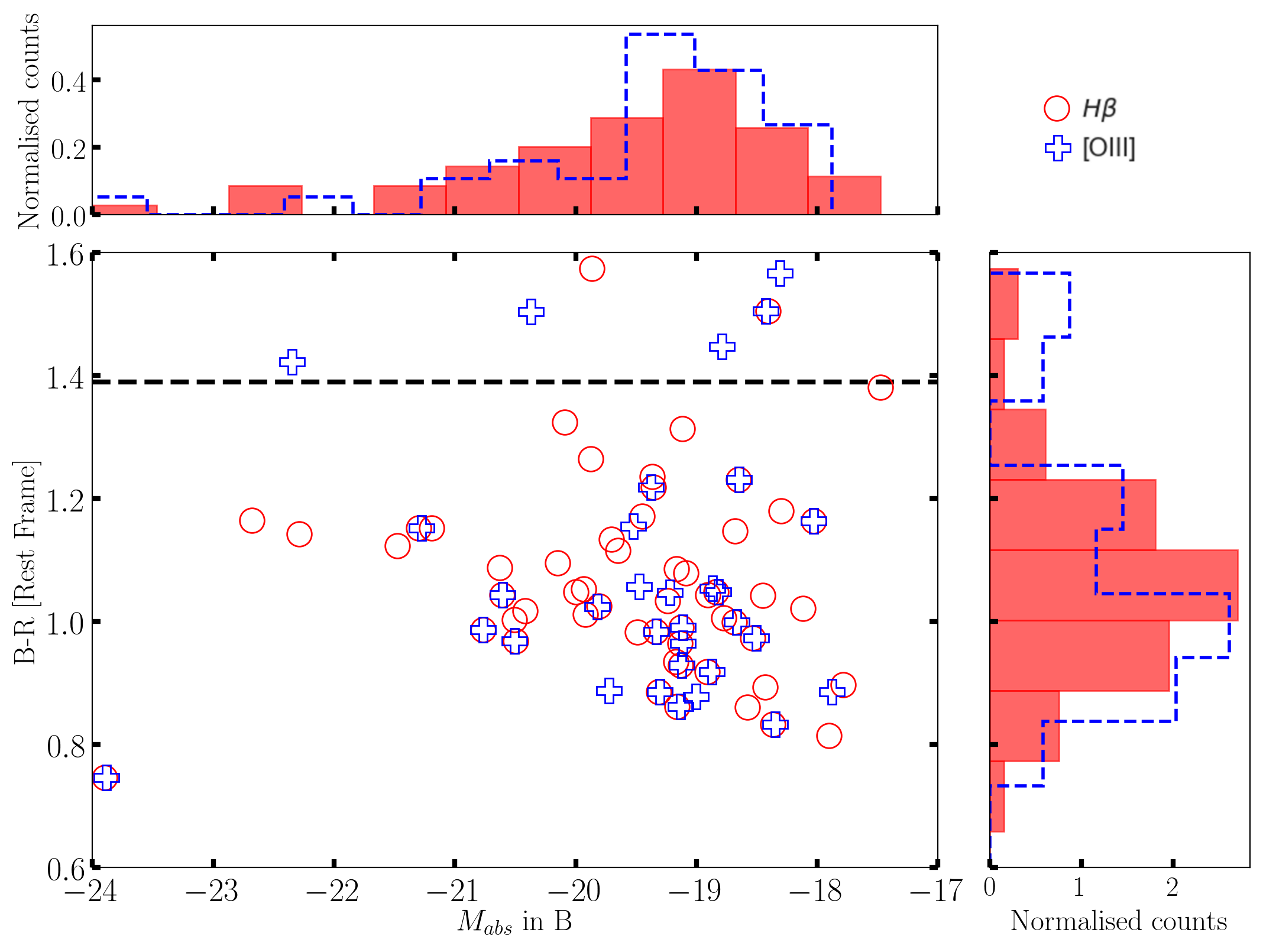}
	 \caption{B\,-\,R rest-frame colour versus absolute magnitude $M_{\rm B}$, for the case of ZwCl0024+1652. The blue crosses represent [O{\sc iii}] emitters while red circles stand for H$\beta$ emitters.} 
	\label{fig:CM_diagram}
\end{figure}

In this section, we discuss some of the galaxy parameters such as colours and stellar masses, in relation to the SF and AGN properties. To begin with, we plotted a colour-magnitude diagram (CMD) for the cluster emitters detected in both [O{\sc iii}] and H$\beta$, with public photometric data gathered from \citet{Treu2003} and \citet{Moran2005}. We applied the k-correction \citep{Blanton2007} to the observed magnitudes to get the rest frame $B\,-\,R$ colour and the absolute magnitude in B. Figure~\ref{fig:CM_diagram} shows the relation between these variables for all [O{\sc iii}] and H$\beta$ emitters. To separate a blue cloud (BC) and red sequence (RS) we introduced a horizontal line at $B\,-\,R\,=\,1.39$ as given by \citet{Sanchez2015}. We determined that only 6 sources (9\%) would belong to the RS while majority (64) of the sources (91\%) remain in the BC region, as expected. The plot includes distributions of both analysed parameters. We do not find any significant difference in the distribution of [O{\sc iii}] and H$\beta$ lines, and most of the sources in both cases (85\% for [O{\sc iii}] and 97\% for H$\beta$) are in the BC, as expected. For the galaxies identified on the CMD, 27 and 13 H$\beta$ emitters are LT and ET, respectively. In the case of [O{\sc iii}] emitters 14 and 8 sources have LT and ET morphologies, respectively. A total of 69 galaxies have spectroscopic classification available, where 42 (60\%) and 27 (40\%) galaxies are SF and AGN, respectively.

\begin{figure}
	\centering
	\includegraphics[width=\columnwidth]{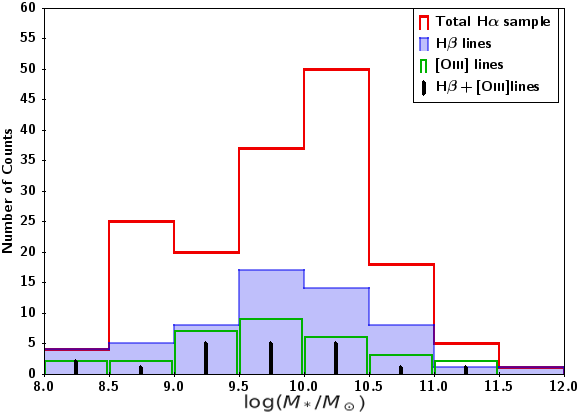}
	\caption{Distribution of the stellar masses. The base sample (H$\alpha$ emitters) is represented with red lines, H$\beta$ lines with blue shaded bars, [O{\sc iii}] emitters with green lines and the black lines represent galaxies with both H$\beta$ and [O{\sc iii}] emission.}
	\label{fig:mass_distribution}
\end{figure}

Stellar mass is another parameter that plays an important role in determining different galaxy parameters such as SFR and sSFR. Previous studies at different redshifts (from the local universe to z\,$\sim$\,2) confirm that stellar mass plays a leading role in determining SFR \citep[e.g.,][]{Koyama2013,Darvish2016,Lagana2018,PiC19}. Using a sample of SDSS galaxies in groups, \citet{Li2019} reported higher SF fractions at all clustercentric distances for lower stellar masses, and higher stellar masses of AGN host galaxies. Moreover, at higher redshifts ($z\,\sim$\,1--2) lower masses of galaxies in clusters are accompanied by enhanced star formation \citep[e.g.,][]{Brodwin2013,Alberts2016} with a peak value at $\mathrm{M}_*\,\simeq\,10^{9.8}\mathrm{M}_\odot$ \citep{Patel2011}. In our case, the stellar mass was computed for each cluster galaxy with {\tt LePhare} code using BC03 templates, as mentioned in Section~\ref{sec:Hb_candidates}. The mass distribution of our base sample (H$\alpha$ sample) and consecutive sub-samples (H$\beta$ and [O{\sc iii}] emitters) is presented in Figure~\ref{fig:mass_distribution} and statistically described in Table~\ref{tab:mass_stat}. From Table~\ref{tab:mass_stat} we can see that the mass distributions for the base sample and consecutive ELG samples are similar, with 50\% of the sample having masses in the range about 9.20\,--\,10.39 in all cases, with median of $\sim$\,9.9. From the histogram (Figure~\ref{fig:mass_distribution}) and the statistics we can see that our consecutive samples cover a complete range of masses from the base H$\alpha$ sample. In the following paragraphs, we discuss different galaxy properties in relation to the effects of stellar mass.

\begin{figure*}
	\centering
	\includegraphics[width=16cm]{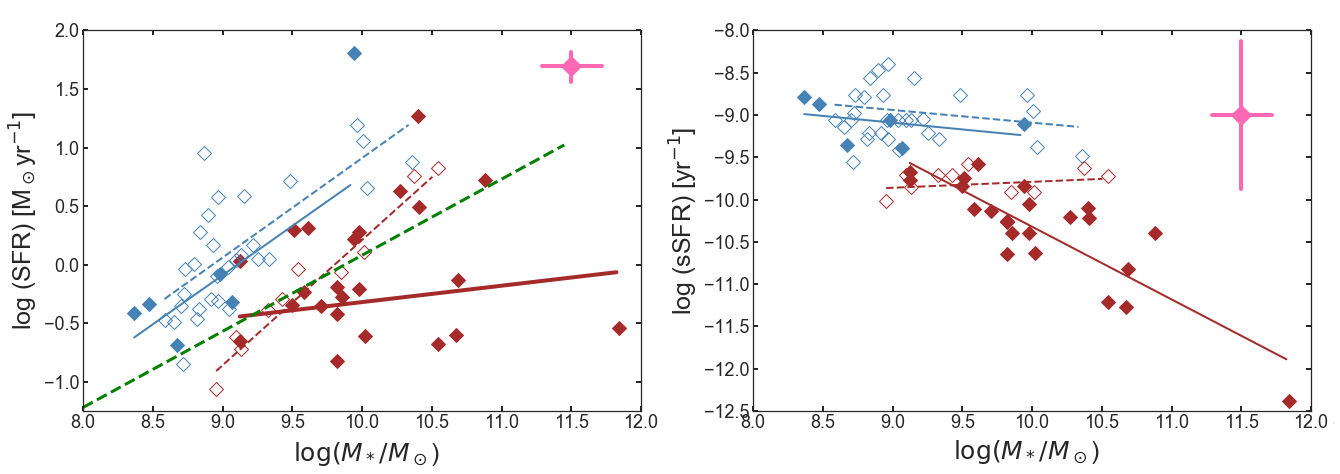}
	\caption{Dependence of the SFR (left) and sSFR (right) on the stellar mass ($\log (M_*/M_\odot)$) for cluster galaxies (filled diamonds with solid linear fittings) and for field galaxies (open diamonds with dashed linear fittings) at z\,$\sim$\,0.4. Active SF and quenching-phase SF galaxies are discriminated with blue and brown colours, respectively. A representative average error bar for cluster SF galaxies is indicated in each plot with pink colour. The dashed green line (left panel) shows the SF main sequence of SDSS galaxies.}
	\label{fig:SFR_mass}
\end{figure*}

\begin{table}
	\centering
	\caption{Statistical description of stellar masses ($\log (M_*/M_\odot)$) of our base H$\alpha$ sample \citep{Sanchez2015} with consecutive identification of H$\beta$ and [O{\sc iii}] emissions.}
	\label{tab:mass_stat}
	\addtolength{\tabcolsep}{-2.5pt}
	\begin{tabular}{lccc} 
		\hline
	        \hline
\textbf{$\log (M_*/M_\odot)$ of}  & \textbf{Q$_1$}  & \textbf{Median} & \textbf{Q$_3$}\\
		\hline
H$\alpha$ base sample & 9.16 & 9.97 & 10.39 \\		
H$\beta$ line emitters & 9.22 & 9.82 & 10.39 \\
$[\rm{OIII}]$ line emitters & 9.22 & 9.82 & 10.39 \\
Both H$\beta$+[O{\sc iii}] line emitters & 9.16 & 9.82 & 10.39 \\
	\hline
	\hline
\end{tabular}
\end{table}
In the first place, we have investigated the behaviour of the SFR and sSFR with stellar mass, as depicted in Figure~\ref{fig:SFR_mass}. Following \cite{Lagana2018}, we further discern between active star-forming galaxies and those in quenching phase, both in field and cluster environments. For this purpose, we used the redshift-dependent sSFR cut, such that sSFR(z)\,=\,10$^{-10}\,\times$\,(1 + z)$^3$, given by \cite{Koyama2013} and the result for cluster galaxies is summarised in Table~\ref{tab:active_quenching}. In Figure~\ref{fig:SFR_mass}, representative average error bars are shown in the plots with average values being 9.37, 0.587 and -10.09, for $\log(M_*/M_\odot)$, \textrm{log}\,(SFR) and \textrm{log}\,(sSFR), respectively. The upper and the lower error limits are determined for each of these parameters. We have observed a rather different behaviour in active SF and quenched  galaxies; on the one hand, the distribution of quenched galaxies extends to higher masses than that of active SF ones, in good agreement with previous results; on the other, we observed that the $\log (M_*/M_\odot)$--$\log(\rm SFR)$ relation of active SF galaxies can be well fitted by a straight line roughly parallel to the main sequence of SDSS SF galaxies (and to that of field galaxies at similar redshift; see Section~\ref{sec:comparison_FC}); however, for quenched galaxies there is a much larger dispersion of values so a linear fit is not appropriate; the behaviour of the sSFR is again remarkable: the sSFR declines mildly with stellar mass for active SF galaxies, but for quenched galaxies this trend is really outstanding, indicating a strong reduction of the SF efficiency with stellar mass. The SFR--M$_*$ correlation in our results, and the anti-correlation of sSFR with M$_*$ conforms with existing results \citep[e.g.,][and references therein]{Vulcani2010,Speagle2014,Vulcani2015,Guglielmo2019}. Concerning the morphologies of SF galaxies, the number of classified galaxies is very low, especially when dealing with ETs. In general, most of LT galaxies are observed to be located slightly below the SDSS main sequence of SF, while the existing small number of ETs has been found above the line.
\begin{table}
	\centering
	\caption{Summary of the placement of our sample SF galaxies into active or quenching phase.}
	\label{tab:active_quenching}
	\addtolength{\tabcolsep}{-2.5pt}
	\begin{tabular}{lcccc} 
		\hline
	        \hline
\textbf{Class}  & \textbf{Cluster}  & \textbf{Field} \\
		\hline
Active SF & 6 (21\%) & 30 (75\%) \\
Quenching SF  & 23 (79\%) & 10 (25\%) \\
Total SF & 29 (100\%) & 40 (100\%) \\
	\hline
	\hline
\end{tabular}
\end{table}

\begin{figure}
\centering
    \includegraphics[width=\columnwidth]{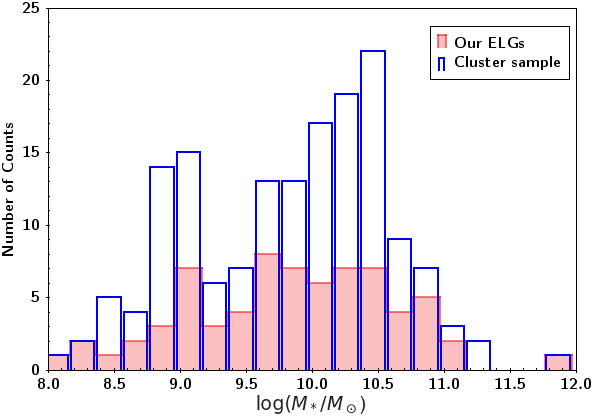}
    \caption{Distribution of our H$\beta$ and/or [O{\sc iii}] ELGs (red filled histograms) and H$\alpha$ cluster sample (blue open histograms).}
    \label{fig:mass_elg}
\end{figure}

Second, we have studied the dependence of the SF and AGN fractions on the stellar mass. The distributions of stellar masses of our sample of ELGs and that of the ancillary H$\alpha$ cluster catalogue follow a similar, bimodal profile with two local maxima at the low- and high-mass ends. The low-mass maximum is placed around  $\log(M_*/M_\odot)$\,=\,9 for both distributions. The high-mass local is placed at $\log(M_*/M_\odot)$\,$\simeq$\,9.8 and $\simeq$\,10.5 for our ELGs and the overall population, respectively. The mass distributions of the overall H$\alpha$ emitting cluster sample and our ELGs are presented in Figure~\ref{fig:mass_elg}.

We finally discuss the effect of stellar masses to the SF and AGN fractions. Within the footprint of the GLACE observations, we have 359 galaxies in the ancillary catalogue of cluster members with derived stellar masses; we have divided the sample into three bins, namely low- ($\log(M_*/M_\odot)$)\,$<$\,9.5), medium- (9.5\,$\leq$\,$\log(M_*/M_\odot)$)\,$<$\,10.5) and high-mass ($\log(M_*/M_\odot)$)\,$\geq$\,10.5) with 109, 135 and 115 galaxies, respectively. The fraction of SF galaxies as a function of the stellar mass is depicted in the right panel of Figure~\ref{fig:SF_frac}. There is a  strong decline in the fraction of SF galaxies in the high-mass bin, confirming that SF is highly suppressed in cluster high-mass galaxies. This agrees with results of \citet{Stasinska2015} whose study implemented the BPT and WHAN diagrams \citep{CidFernandes2011}. For high mass bin, the small fraction of SF galaxies in our cluster could be because the galaxies in this bin could have formed almost all the stars at redshifts higher than z\,$\sim$\,0.4. Finally, the fraction of AGN galaxies as a function of the stellar mass is depicted in the right panel of Figure~\ref{fig:AGN_frac}. There is a slight decline of the fraction of AGNs both at the low- and high- mass bins, although not very significant.

\subsection{The phase--space relation and the local density effects}
\label{sec:sigma5}
\begin{figure}
\centering
    \includegraphics[width=\columnwidth]{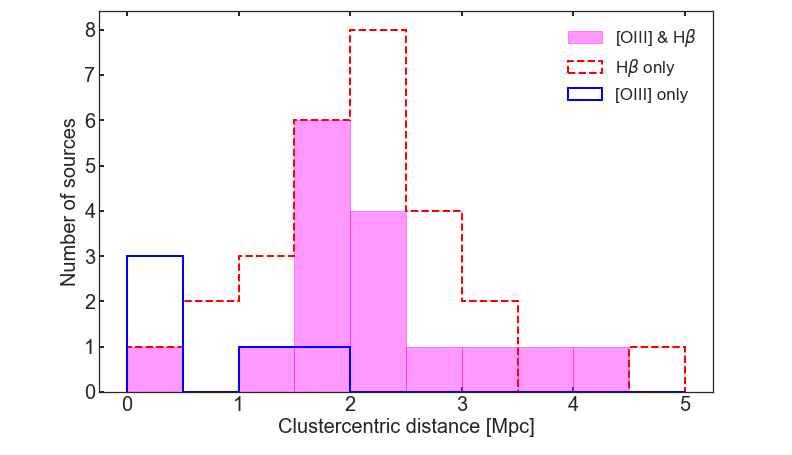}
    \caption{Distribution of [O{\sc iii}] and H$\beta$ emission lines in ZwCl0024+1652 cluster with clustercentric distance, for members of the main cluster structure.}
    \label{fig:lines_distribution}
\end{figure}

We adopted the centre coordinates of the cluster as RA$_{\rm c}$\,=\,0\,$^{\rm h}$26\,$^{\rm m}$45.9\,$^{\rm s}$ and DEC$_c$\,=\,17$^{\rm d}$\,9$^{\rm m}$41.1\,$^{\rm s}$, based on the work of \citet{Diaferio2005}. Taking into account the distance of the cluster of 1500\,Mpc and following the procedure outlined in \citet{Amado2019}, we computed the clustercentric distance of each identified ELG. The radial velocity was also computed and the phase--space diagram was plotted to identify the distribution of the sources (both [O{\sc iii}] and H$\beta$ emitters) as shown in Figure~\ref{fig:velocity}. We limited the sample to the main structure of the cluster (0.385 $<$ z $<$ 0.41) and excluded the BLAGN galaxies as described in Section \ref{sect:BPT_analysis}. The resulting distributions of H$\beta$ and [O{\sc iii}] line emitters in clustercentric distance are given in Figure~\ref{fig:lines_distribution}. Moreover, the phase--space diagram with two plots (for H$\beta$ and [O{\sc iii}]) showing the line of site velocity versus the projected clustercentric distance is introduced in Section~\ref{sec:Results}, Figure~\ref{fig:velocity}. Combining the relationships observed from both informations (clustercentric distance distribution and phase--space diagrams) one can comment on the location of member galaxies inside the cluster \citep[e.g.,][]{Jaffe2015}. From the plots in Figures~\ref{fig:velocity} and~\ref{fig:lines_distribution} it can be seen that [O{\sc iii}] emitters are relatively closer to the cluster centre than H$\beta$ line emitters despite all our emitters are with in r$_{\rm vir}$ of our cluster. One of these sources has one of the highest dust extinctions (A$_{\rm H\alpha}=2.43$ mag). For the remaining three [O{\sc iii}]-only SF galaxies we estimated extinctions using the H$\beta$ limiting flux given in Table \ref{tab:flux}, obtaining A$_{\rm H\alpha}$ values between 1 and 2 mag. This fact could explain the lack of H$\beta$ detection for those objects. On the other hand, we verify that about half of [O{\sc iii}]- and H$\beta$-only emitters are located in intermediate clustercentric distances. The lack of  [O{\sc iii}]-only emitters at large clustercentric distances is intriguing, however we would need larger samples to confirm this and to understand the possible cause.

For discussion related to environment concentrating in the main cluster structure, we found 35 SF galaxies and only 13 AGNs within the clustercentric distance range of 0\,-\,5\,Mpc. Despite the small number statistics, we found that the AGN hosts belonging to the main structure of the cluster are concentrated inside $\sim$\,1.5 virial radii ($\sim$\,3\,\textrm{Mpc}). This contrasts to the results given by \citet{Sobral16}, who do not found any strong dependence of AGN fraction on environment for the case of Cl 0939+4713 at z\,$\sim$\,0.4. To support our results, we presented a phase--space relation in Figure~\ref{fig:velocity} clearly representing the distribution of SF and AGN galaxies with [O{\sc iii}] and H$\beta$ emission lines. Our findings are in line with previous results at intermediate redshifts reporting that the fraction of AGN increases with clustercentric distance, confirming the environment dependence on AGN activity \citep[e.g.,][]{Argudo-Fernandez2018, Koulouridis2019, Hora2020}. On the other hand, SF galaxies are distributed across all measured clustercentric distances, but they are preferentially located between 0.5 and 1.5 virial radii, with the bulk of SF galaxies exhibiting the lowest extinction values.

\begin{figure*}
	\centering
	\includegraphics[width=16cm]{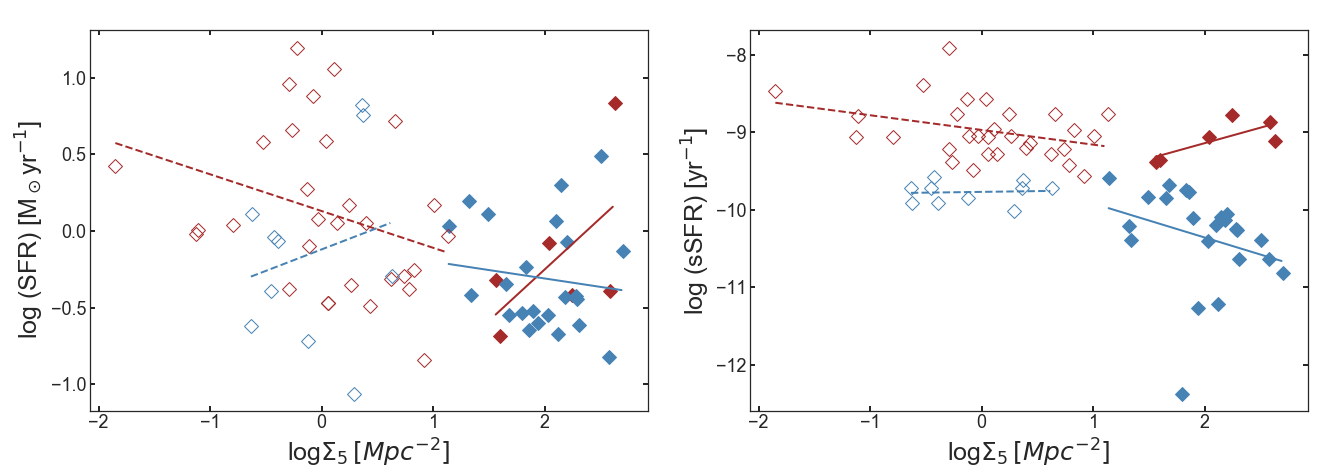}
	\caption{Dependence of the SFR (left) and sSFR (right) on the local density ($\Sigma_5$) for cluster galaxies (filled diamonds with solid linear fittings) and for field galaxies (open diamonds with dashed linear fittings) at z\,$\sim$\,0.4. Active SF and quenching-phase SF galaxies, separated as described in Section~\ref{sec:mass}, are discriminated with brown and blue colours, respectively.} 
	\label{fig:SFR_sigma}
\end{figure*}

We computed a valid local density estimation for 72 from 73 sources of the main cluster sample using the algorithm from \cite{Dre85}. We also computed the local surface density with the same approach for VVDS-Deep field SF galaxies selected as described in Section~\ref{sec:field_data}. Figure~\ref{fig:SFR_sigma} describes the relationship of SFR and sSFR to local density where it is shown separately for active as well as quenching phase SF galaxies.
Previous studies suggested that, both in the local universe and at higher redshift, regions with higher projected local density are generally characterised by smaller fractions of galaxies with ongoing star formation \citep[e.g.,][]{Pimbblet2002,Gomez2003,Kauffmann2004}. On the other hand, different works have shown that SF could be triggered and/or quenched by the nature of environment in which the galaxy is located, as can be described through relations between different parameters (such as colour, morphology, stellar mass and SFR) and environment \citep[e.g.,][and references therein]{Peng2010ApJ,Peng2012,Tomczak2019}. \citet{Poggianti2008} used the [O{\sc ii}] emission line as a proxy of the SFR of galaxies in clusters in the range 0.4\,$<$\,z\,$<$\,0.8,  showing that the SFR is lower in denser environments \citep[see also][]{Rudnick2017}. Recently, \citet{Lagana2018} derived the SFR of cluster galaxies at 0.4\,<\,z\,<\,0.9 from broadband SED fitting, and found no clear  dependence between either SFR or sSFR and local environment. \citet{PiC13} also studied the dependence of the  SF activity derived from IR luminosity with local environment in a young galaxy cluster at z\,$\sim$\,0.87, finding that the average SFR and sSFR are roughly independent of the local environment. However, the fraction of SF galaxies varies with the local density, being enhanced in the intermediate density regions and dropping in the highest local density ones. The same behaviour is observed in both high- and low-mass galaxies. From the projected phase space diagrams in Figures~\ref{fig:velocity} and~\ref{fig:velocity_morpho} we have determined that the most central region of the cluster ($r\,\lesssim\,0.3Mpc$) is almost devoid of ELGs, leading to a conclusion that the star formation and the AGN activity could be completely quenched. We can also deduce that almost all ELGs are located in the main cluster structure (structure ``A'').
In a recent work, \citet{PiC19} further studied the population of SF galaxies within a sample of IR-selected galaxy clusters in the range  0.3\,$\leq$\,z\,$\leq$\,1.1 by means of broadband SED fitting, finding a decrease of the fraction of SF galaxies from field towards cluster cores, strongly dependent on stellar mass and redshift.

 The relations among SFR and galaxy environment have been studied by many authors in our redshift range \citep[e.g.,][]{Vulcani2010,Stroe2015}. In our current work, we have explored the SF activity (as given by the SFR and sSFR) for the purely SF galaxies (after removing AGN sources) as a function of the environment, as depicted in Figure~\ref{fig:SFR_sigma}. No clear correlation between the SFR and local density is observed for either active SF or quenched cluster galaxies, in line with previous results. However, for the sSFR we found a remarkable behaviour: the sSFR seems to be correlated with local density for active SF galaxies, and anti-correlated for quenched ones. While the former relation is mild and could be impacted by the small size of the sub-sample (5 galaxies), the latter seems to be stronger and supported by the larger size of the set of quenched galaxies. Even though there is a mild, highly dispersed correlation between stellar mass and local density that can somehow contaminate the result, this finding suggests that star formation efficiency declines with the increase of the local density of the environment. Moreover, this trend, but with a slope about 1.4 times shallower, has been also reported by \citet{Chung2011} in a comprehensive study of a sample of 69 clusters at z\,$<$\,0.1. Hence, this suggest a possible evolution of the quenching strength of SF galaxies, which evidently requires further studies. Moreover, morphologies of the SF galaxies have been observed against active and quenching SF phases. In active SF phase, 33\% of galaxies with known morphologies are ET while 67\% are LT. Similar has been found for quenching SF phase, where 37\% and 63\% of sources are ET and LT, respectively. Our results show that getting into quenching phase from active SF phase, the ET and LT proportions look nearly the same (with slight increase of ET fraction with a slight decrease in LT proportion); not exactly conforming with previous results \citep[e.g.,][]{Kelkar2019} showing relatively suppressed star formation in older stellar populations. That could be because of small statistics in our current work.
  \begin{figure*}
	\centering
	\includegraphics[width=9cm]{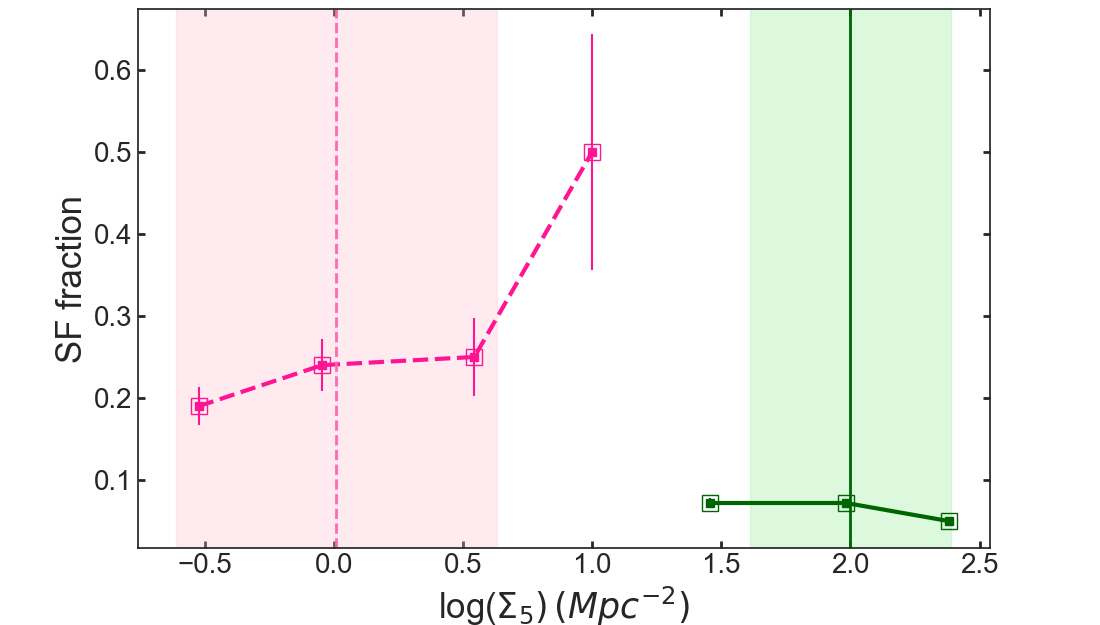}
	\hspace{-1cm}
	\includegraphics[width=9cm]{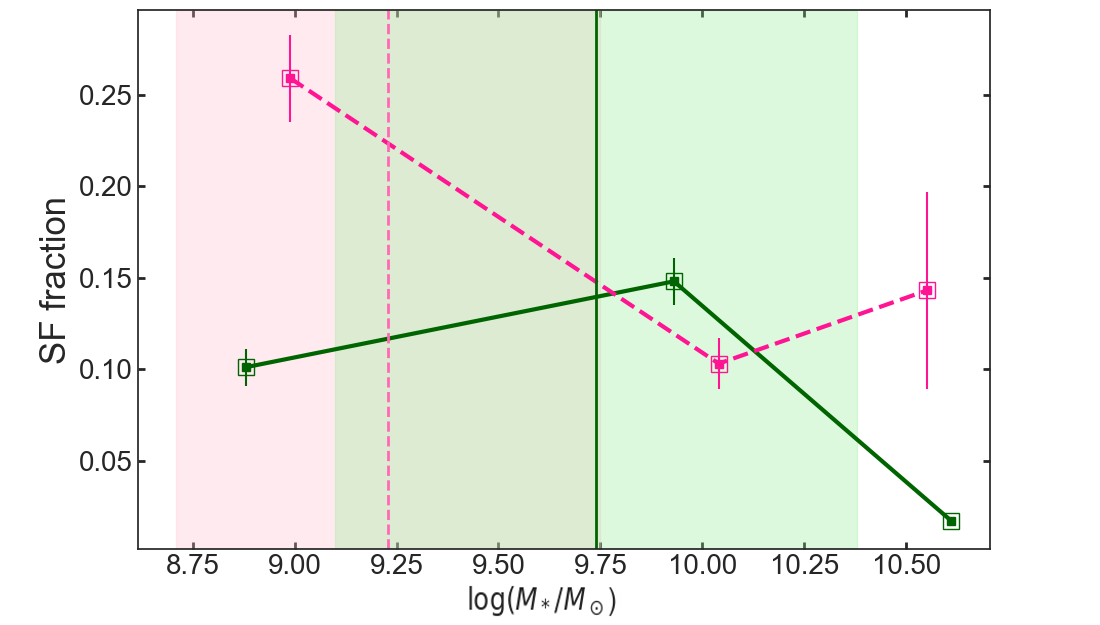}	
	\caption{Variation of the fraction of SF galaxies versus local density (left panel) and stellar mass (right panel). The solid green lines show cluster galaxies and the pink dashed lines show field galaxies at z\,$\sim$\,0.4. In the right panel, the sample is divided into three mass bins with low: $\log(M_*/M_\odot)<9.5$, medium: 9.5\,$\log(M_*/M_\odot)<10.5$ and high: $\log(M_*/M_\odot)>10.5$. Error bars correspond to Poissonian statistics  (see the text). Vertical lines represent the corresponding mean values with a light shaded standard deviation.}
	\label{fig:SF_frac}
\end{figure*}

\begin{figure*}
	\centering
	\includegraphics[width=9cm]{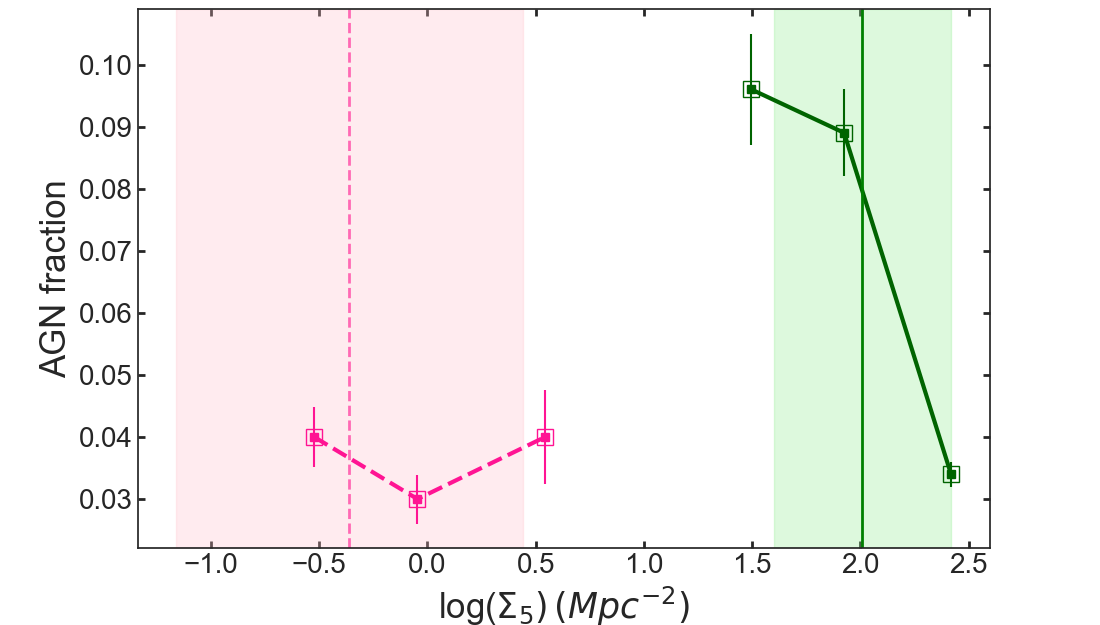}
	\hspace{-1cm}
	\includegraphics[width=9cm]{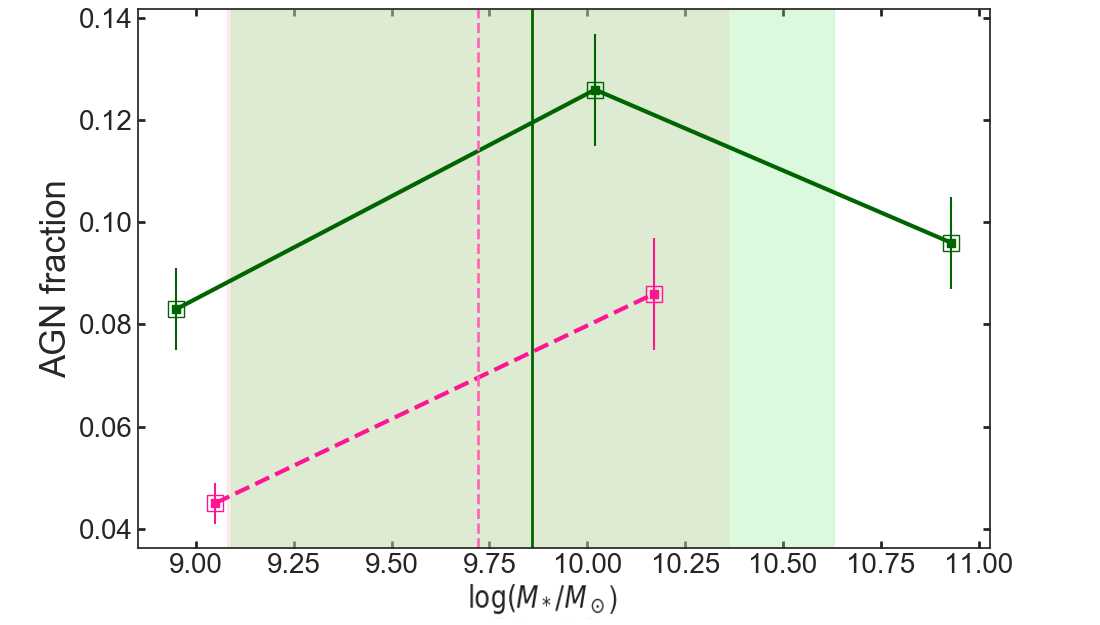}	
	\caption{Variation of the fraction of AGN galaxies versus local density (left panel) and stellar mass (right panel). The solid green lines show cluster galaxies and the pink dashed lines show field galaxies at z\,$\sim$\,0.4. In the right panel, the sample is divided into three mass bins with low: $\log(M_*/M_\odot)<9.5$, medium: 9.5\,$\leq\,\log(M_*/M_\odot)<10.5$ and high: $\log(M_*/M_\odot)>10.5$. Error bars correspond to Poissonian statistics  (see the text). Vertical lines represent the corresponding mean values with a light shaded standard deviation.}
	\label{fig:AGN_frac}
\end{figure*}

Finally, in order to investigate the dependence of the fraction of SF galaxies on the environment, we have determined the SF fractions as a function of the local density $\Sigma_5$ as shown in the left panel of Figure~\ref{fig:SF_frac}. As in \citet{PiC13}, we have followed the prescription of \citet{Koy2008}, who defined the intermediate density environment as a relative narrow range of local density where the optical colour distribution starts to change dramatically. According to their definition, the low-density environment corresponds to $\log \Sigma_5 < 1.65$, the medium-density range to $1.65 \leq \log \Sigma_5 < 2.15$ and finally the high-density range to $\log \Sigma_5 \geq 2.15$. \citet{PiC13} demonstrated that the derived parameters are robust against small random variations of the boundaries, and therefore the selection does not condition the results of the environmental analysis. The total number of objects from the ancillary cluster catalogue within the sky area of our survey and with $R_{\rm AB}$\,$<$\,23 are 86, 139 and 222 in the low-, medium- and high-density bins, respectively.
The fractions were computed as the ratio of the number of SF galaxies in each bin to the total number of member galaxies in the same bin. The uncertainty in each fraction was computed as the ratio of the fraction in the bin to the square root of the total number of galaxies in that bin. 
It can be observed that the fraction of SF galaxies smoothly declines across the low- and medium-density bins, dropping towards the high-density region. Hence, we cannot support the claim of \citet{PiC13} that the intermediate density environment favours the enhancement of the SF activity, although it does not inhibit it (or quenching and triggering effects compensate). Authors of \citet{PiC13} gave results based on low $\log(M_*/M_\odot)<10.5$ and high ($\log(M_*/M_\odot)\,\geq\,10.5$) mass bins, concluding that high density regions have scarcity of low mass SF galaxies. Moreover, they reported that high sSFR values are observed in low density regions, where high mass systems characterized by low sSFR dominate the high density environment. While examining our results in high and low mass bins as in \citet{PiC13} we observed significant number of low mass SF galaxies. In addition to this, we found that low mass SF galaxies posses high sSFR independent of environment while high mass SF galaxies in high density region have low sSFR values. Using the same bins and a similar procedure, we computed the AGN fraction, as shown in the left panel of Figure~\ref{fig:AGN_frac}, finding a similar trend as for SF galaxies, i.e. declining throughout the low- and medium-density environments and dropping sharply at higher densities. This is in line with previous results \citep[e.g.,][]{Li2019}.  The mean local densities of SF galaxies and AGN are ${\rm log}\,(\Sigma_5)$ [{\rm Mpc}$^{-2}$]\,=\,2.00\,$\pm$\,0.39 and 1.82\,$\pm$\,0.34, respectively. 

\subsection{Cluster-field comparison at $z\,\sim\,0.4$}
\label{sec:comparison_FC}
Both the field and cluster galaxies are consistently selected in a similar approach. Figure \ref{fig:SFR_sigma} shows that both are quite smoothly distributed in a local density range covering  $\sim$\,5 decades. We are confident that GLACE data also reasonably samples the lower-density outskirts of this cluster out to $\sim2r_{vir}\approx\,4$\,\rm{Mpc} (with OSIRIS TF giving a circular FOV of 4\,\rm{arcmin} radius), despite the angular coverage of our observations is not exhaustive. We assume that VVDS data is fairly representative of field galaxies at the cosmic epoch studied here.

\begin{figure}
	\centering
	\includegraphics[width=\columnwidth]{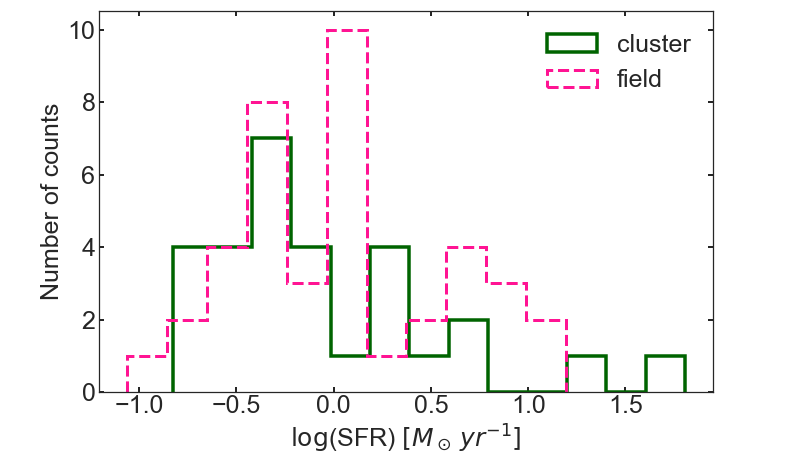}
	\caption{Histogram representing the distribution of SFR for SF galaxies in ZwCl0024+1652 (green solid) and field (dashed pink) lines. The mean value of {\rm log}(SFR)\, [{\rm M}$_\odot$\,{\rm yr}$^{-1}$]\,=\,-0.06\,$\pm$\,0.60 for cluster members while the mean {\rm log}(SFR)\, [{\rm M}$_\odot$\,{\rm yr}$^{-1}$]\,=0.05\,$\pm$\,0.054.}
	\label{fig:SFR_comp}
\end{figure}

Evidence from large-area surveys \citep[e.g.,][]{Gomez2003,Bongiovanni2005} shows that the volume-averaged SFR monotonically decreases with higher environmental densities. In particular, higher average SFRs are found in the field at different redshifts from z$\sim$1 when compared to galaxy cluster environments \citep[e.g.,][]{Linden2010,Muzzin2012,Paulino-Afonso2018,Hwang2019}. At higher redshifts, SFRs of cluster galaxies showed to be greater than in the case of field galaxies \citep[e.g.,][]{Tran2010,Alberts2014}. In Figure~\ref{fig:SFR_comp} we show a comparison of SFRs of the ZwCl0024+1652 cluster and field galaxies from VVDS at the same cosmic epoch. A global difference of $\sim$0.1 dex between both sets, with higher SFRs obtained for field galaxies than cluster members, according to the results referred above. 

Regarding the relation between the SFR and the local density ($\Sigma_5$), it is not clear whether exists a significant change of the SFR with this parameter for both, cluster and field environments, independently from their evolutionary phase, as shown in the left panel of Figure \ref{fig:SFR_sigma}. In contrast, 
the environmental dependences of the sSFR are clearer, as shown in the right panel of Figure \ref{fig:SFR_sigma}. In fact, the ratio of field active SFGs to quenching (or transient) phase ones in the sSFR-$\Sigma_5$ relation is turned upside down with respect to that of the cluster. At redshift $z\,\sim\,0.4$, we found $\sim$\,21\,\% ($\sim$\,79\,\%) and $\sim$\,75\,\% ($\sim$\,25\,\%) of active SFGs (quenching phase) cluster and field galaxies, respectively. Hence, while in low density environments there is a relatively low fraction of transient SF galaxies without a measurable dependence on the local density, in the case of this cluster we found a remarkable environment-depending SF quenching, as already mentioned in Section \ref{sec:sigma5}. This finding agrees with previous studies carried out at different redshifts (e.g.) \citet{Cucciati2017}, who used VIMOS Public Extragalactic Redshift Survey (VIPERS) to analyse environmental effect on the evolution of galaxies at $0.5\,\leq\,z\,\leq\,0.9$, \citet{Baldry2006} for environment studies carried out in the local universe, and \citet{Cucciati2010} for intermediate redshift studies, but in galaxy groups. All of these studies have found in common that the ratio of active to quenching SF galaxies is smaller in higher density environments. The mean fraction of the field SFGs is comparable with that of cluster in the low-density regime, as illustrated in Figure~\ref{fig:SF_frac}.

About the relationships with stellar mass, there is a well-established relation between this parameter and SF phenomena, both in field and cluster \citep[e.g.,][]{Noeske2007,Peng2010ApJ,LaraLopez2010,Speagle2014}. In general, the SFR--M$_*$ relation has a bivariate behaviour depending on the SF activity. The active SF galaxies populate the main sequence, whereas the quenched (e.g. post-starbursts) ones tend to be grouped horizontally in this diagram.

In Figure~\ref{fig:SFR_mass} (left panel) we show the dependence of SFR on stellar mass both for field and cluster SFGs. Accordingly, {\rm log}\,({\rm SFR}) and $\log (M_*/M_\odot)$ are positively correlated with slopes 0.12\,/\,0.64, respectively for cluster/field SF galaxies. As a reference, we have represented the main sequence (MS) traced by low-redshift SFGs from SDSS \citep{Brinchmann2004}. The active SF galaxies of the cluster tends to populate a region that roughly corresponds to the MS, especially for large masses, while active SF and a few galaxies in quenching phase of the field reside above it. 

From the definition given above, the sSFR traces the contribution of the star-formation processes to the mass growth of the galaxies. In general, low-mass galaxies tend to have higher sSFR that their high-mass counterparts \citep[see][and references therein]{Vulcani2010}. This is just one of consequences of the 'downsizing' scenario, which favours the formation of massive galaxies in earlier epochs and shorter time-scales and the opposite situation for low-mass galaxies. From the Figure~\ref{fig:SFR_mass} (right panel) it can be seen that a general inverse relation is observed between the {\rm log}\,({\rm sSFR}) and $\log (M_*/M_\odot)$, both field and cluster, in agreement with this scenario but at different mass scales and the same cosmic epoch. This fact suggest that the cluster environments favours a rapid transition from active SF to quenched phase compared with the field. In other words, the high-density environments at intermediate redshifts are real laboratories for testing the properties of low-redshift galaxies of high mass.

Finally, despite the relatively small statistics in the field sample, we can clearly see that irrespective of the local density values the SF fraction in the core of the cluster is around a half of that in the field environment. On the other hand, the trend of the SF fractions of field galaxies as a function of stellar mass is opposite to that of cluster galaxies. This suggests an accelerated evolution of cluster galaxies with respect to field ones. In contrast, the AGN fractions as a function of M$_*$ have the same trend.

\section{Conclusions}
\label{sec:Conclusions}

In this work we are demonstrating the capabilities of tunable filters (TF) to study ELG populations in galaxy clusters. We focused our study on ZwCl0024+1652 cluster at an intermediate redshift of $z\,\sim\,0.4$, observed using the GTC/OSIRIS instrument under the GLACE survey. This paper describes TF observations, data reduction, sample selection, and analysis of [O{\sc iii}] and H$\beta$ emission lines. We compared the obtained fluxes and luminosities of [O{\sc iii}] and H$\beta$ lines with H$\alpha$ and [N{\sc ii}] lines measured in \citet{Sanchez2015}. We studied different properties of [O{\sc iii}] and H$\beta$ emitters in terms of their clustercentric distance, morphology, star formation, AGN activity, local density, and stellar mass. Finally, we compared the properties of our cluster galaxies with the field galaxies at the same redshift using data from the VVDS-Deep survey. Our main findings are:  
\begin{itemize}
\item Out of 174 previously confirmed H$\alpha$ emitters \citep{Sanchez2015}, 35 ($\sim20\%$) and 59 ($\sim34\%$) galaxies show [O{\sc iii}] and H$\beta$ emission lines, respectively, out to the H$\alpha$ flux of 2.43\,$\times$\,10$^{-15}$\,{\rm erg}\,{\sc s}$^{-1}$\,{\rm cm}\,$^{-2}$. Our TF redshifts computed from the position of H$\beta$ and [O{\sc iii}] lines using the pseudo-spectra have been compared with previous results of \citet{Sanchez2015}, finding a good correlation between all  measurements.   
\item For 20 galaxies we were able to reproduce the BPT-NII diagram finding 8 (40\,\%) SF galaxies, 11 (55\,\%) composites, and 1 (5\,\%) LINER. In terms of morphology, we found higher fraction of LT ($\sim$\,67\,\%) sources in comparison to ETs ($\sim$\,33\,\%), showing that ELG are predominantly late-types with SF and/or AGN activity going on.
\item From the projected phase--space and the clustercentric distance analysis, we determined that core of this cluster is relatively free of ET galaxies with emission lines up to the clustercentric distance of $\sim$\,500\,{\rm kpc}.
\item We found that both SF galaxies and AGN follow the local density-clustercentric distance anti-correlation, and we confirmed using TF observations that cores of regular galaxy clusters are poor with SF galaxies and AGN up to the clustercentric distance of $\sim$\,1.3\,Mpc. This is in line with most of previous studies where SF and AGN activity has been detected using other than TF data. 
\item The sSFR seems to be anti-correlated with local density for quenched galaxies, suggesting that the star formation efficiency declines with the increase of local density of the environment.
\item The fraction of SF galaxies and AGNs remain roughly constant across the low- and medium-density environments, dropping sharply towards high-density environments.
\item The distribution of quenched galaxies extends to higher masses than that of active SF ones. The $\log(M_*/M_\odot)$--$\log(\rm SFR)$ relation of active SF galaxies can be well fitted by a straight line roughly parallel to the main sequence of SF galaxies.
\item The sSFR declines mildly with stellar mass for active SF galaxies, but for quenched galaxies this trend is really outstanding, indicating a strong reduction of the SF efficiency with stellar mass.
\item There is a  strong decline in the fraction of SF galaxies in the high stellar mass range, confirming that SF is highly suppressed in cluster high-mass galaxies. However, no remarkable trend with respect to stellar mass is observed for AGNs. 
\item Cluster-field comparison in terms of star formation processes suggest an accelerated evolution of galaxies in cluster with respect to field.
 \end{itemize}
This work provides an important contribution to previous studies regarding the properties of emission line galaxies in clusters, and shows the importance and strength of TF data to study galaxy clusters even at higher redshifts. An example of catalogue with properties of detected emission line galaxies is given in Appendix of this paper, while the full catalogue will be available in the electronic version.

\section*{Acknowledgements}
We acknowledge the anonymous referee for helpful comments and valuable suggestions that significantly contributed to improve the paper. We thank the Ethiopian Space Science and Technology Institute (ESSTI) under the Ethiopian Ministry of Innovation and Technology (MOIT) for all 
the financial and technical supports. ZBA specially acknowledges Instituto de Radioastronom\'ia Milim\'etrica (IRAM), Spain, for giving financial support and working space, and also extend thanks to Instituto de Astrof\'isica de Andaluc\'ia (IAA-CSIC), Spain, for providing a working space for part of the work. ZBA further extends special thanks to Joint ALMA Observatory (JAO) visitors' program for the financial and technical support during the data reduction. Moreover, ZBA acknowledges Kotebe Metropolitan University for granting a study leave and giving material supports. MP acknowledges support from the Spanish MCIU under project AYA2016-76682-C3-1-P, and from the State Agency for Research of the Spanish MCIU through the ``Center of Excellence Severo Ochoa'' award for the Instituto de Astrof\'isica de Andaluc\'ia (SEV-2017-0709).  MCS and APG are funded by Spanish State Research Agency grant MDM-2017- 0737 (Unidad de Excelencia Mar\'ia de Maeztu CAB).
This work was supported by the Spanish Ministry of Economy and Competitiveness (MINECO) under the grants AYA2017-88007-C3-2-P.
In this work, we made use of Virtual Observatory Tool for OPerations on Catalogues And Tables (TOPCAT), Python scientific codes and IRAF. IRAF is distributed by the National Optical Astronomy Observatories, which are operated by the Association of Universities for Research in Astronomy, Inc., under cooperative agreement with the National Science Foundation. We also used ACS/HST data based on observations made with the NASA/ESA HST, and obtained from the Hubble Legacy Archive, which is a collaboration between the Space Telescope Science Institute (STScI/NASA), the Space Telescope European Coordinating Facility (ST-ECF/ESA) and the Canadian Astronomy Data centre (CADC/NRC/CSA). 
\section*{Data Availability}
The descriptions of the data underlying this article are available in the article and an online table as supplementary material.



\newpage
\appendix

\section{Catalogue of [O{\sc iii}] and/or H$\beta$ emitters in ZwCl0024+1652 galaxy cluster}

A catalogue of emission line galaxies in ZwCl0024+1652 cluster is presented in this paper. Table~\ref{tab:Catalogue} gives only an example, while the full catalogue will be given in the electronic version of paper. Properties of [O{\sc iii}] and H$\beta$ emission lines are given for 35 and 59 galaxies, respectively. We added also the data of [NII] and H$\alpha$ emitters studied in \cite{Sanchez2015}. The entire catalogue gives the following properties of 73 galaxies:\\
\textbf{Column 1} ... H$\alpha$ line ID number with subscript O for offset and C for centre position;\\
\textbf{Column 2} ... [O{\sc iii}] line ID number with subscript O for offset and C for centre position; =\,-99 if not available;\\
\textbf{Column 3} ... H$\beta$ line ID number with subscript O for offset and C for centre position; =\,-99 if not available;\\
\textbf{Column 4} ... HST ID number; =\,-99. if not available;\\
\textbf{Column 5} ... Right Ascension in decimal degrees (J2000);\\
\textbf{Column 6} ... Declination in decimal degrees (J2000);\\
\textbf{Column 7} ... Clustercentric distance in {\rm Mpc};\\
\textbf{Column 8} ... TF redshift measured with [O{\sc iii}] line; =\,99. if not available;\\
\textbf{Column 9} ... TF redshift measured with H$\beta$ line; =\,99. if not available;\\
\textbf{Column 10} ... TF redshift measured with H$\alpha$ line; =\,99. if not available;\\
\textbf{Column 11} ... [O{\sc iii}] flux in $\times\,10^{-16}\,\mathrm{erg}\,\mathrm {s}^{-1}\,\mathrm {cm}^{-2}$; =\,-99. if not available;\\
\textbf{Column 12} ... H$\beta$ flux in $\times\,10^{-16}\,{\mathrm erg}\,{\mathrm s}^{-1}\,{\mathrm cm}^{-2}$; =\,-99. if not available;\\
\textbf{Column 13} ... H$\alpha$ flux in $\times\,10^{-16}\,{\mathrm erg}\,{\mathrm s}^{-1}\,{\mathrm cm}^{-2}$; =\,-99. if not available;\\
\textbf{Column 14} ... [N{\sc ii}] flux in $\times\,10^{-16}\,{\mathrm erg}\,{\mathrm s}^{-1}\,{\mathrm cm}^{-2}$; =\,-99. if not available;\\
\textbf{Column 15} ... H$\beta$ equivalent width in \AA; =\,-99. if not available;\\
\textbf{Column 16} ... List of emission lines detected;\\
\textbf{Column 17} ... Base 10 logarithm of stellar mass, $\log (M_*/M_\odot)$; =\,-99. if not available;\\
\textbf{Column 18} ... Emission line galaxy (ELG) type based on \citet{Sanchez2015};\\
\textbf{Column 19} ... BPT-NII classification; =\,-99.9 if not available;\\
\textbf{Column 20} ... Broad morphological class, early-type (ET) or late-type (LT); 99.9 if not available; \\
\textbf{Column 21} ... ource of the morphological class (A\,=\,\citet{Treu2003}, B\,=\,\citet{Moran2007}, C\,=\,\citet{Amado2019}); 99.9 if not available;\\

	\begin{table*}
	\center
	\caption{Example of the catalogue describing [O{\sc iii}] and H$\beta$ emitters in ZwCl0024+1652 cluster (Full catalogue is available online).}
	\label{tab:Catalogue}
	\begin{tabular}{llllll} 
		\hline\hline
		H$\alpha$\_ID & [O{\sc iii}]\_ID & H$\beta$\_ID & HST\_ID & RA (deg) & DEC (deg) \\ 
		R (Mpc) & z\_[O{\sc iii}] & z\_H$\beta$ & z\_H$\alpha$ & $f_{[O{\sc iii}]}$ & $f_{H\beta}$ \\ 
		$f_{H\alpha}$ & $f_{[N{\sc ii}]}$ & EW\_H$\beta$ & Line\_detected & $\log (M_*/M_\odot)$ & ELG\_Type \\
		BPT\_class & Morpho\_class & Morpho\_source \\
				\hline\hline
		1015\_O & -99 & 1009\_O & -99 & 6.6232 & 17.2470 \\   
		2.8131 & 99. & 0.3955 & 0.3937 & -99. & 1.4467 \\
		4.6594 & 1.3600 & 5.1800 & H$\beta$,H$\alpha$,[N{\sc ii}] & 9.9446 & SF \\
		 99.9 & 99.9 & 99.9  \\
				\hline
		1039\_O & -99 & 946\_O & 136 & 6.5373 & 17.2530 \\
		4.5367 & 99. & 0.3959 & 0.3961 & -99. & 0.7424 \\
		2.8236 & 2.2306 & 2.1550 & H$\beta$,H$\alpha$,[N{\sc ii}] & 10.8850 & BLAGN   \\
		99.9 & ET & A,B  \\
				\hline
		1057\_O & 1056\_O & -99 & -99 & 6.5649 & 17.2297 \\
		1.8003 & 0.3922 & 99. & 0.3934 & 0.2523 & -99. \\
		1.3030 & 0.4714 & -99. & [O{\sc iii}],H$\alpha$,[N{\sc ii}] & 9.9831 & SF \\
		99.9 & 99.9 & 99.9 \\
				\hline
		1097\_O & 1068\_O & 1203\_O & -99 & 6.6542 & 17.2282 \\
		3.6278 & 0.3957 & 0.3954 & 0.3947 & 0.2044 & 0.2124 \\
		0.8933 & 0.4282 & 4.3030 & H$\beta$,[O{\sc iii}],H$\alpha$,[N{\sc ii}] & 9.2247 & SF \\
		composite & 99.9 & 99.9 \\
		   \hline
		1157\_O & 1134\_O & 1306\_O & 109 & 6.5377 & 17.2212 \\
		4.1462 & 0.4057 & 0.4039 & 0.4042 & 1.2396 & 1.3597 \\
		5.3748 & 0.9067 & 3.976 & H$\beta$,[O{\sc iii}],H$\alpha$,[N{\sc ii}] & 10.4082 & SF \\
		SF & LT & A \\
		   \hline
		1173\_O & 810\_O & 902\_O & -99 & 6.6182 & 17.2600 \\
		3.1621 & 0.3876 & 0.3917 & 0.3901 & 0.6628 & 0.6482 \\
		1.7725 & 0.5727 & 5.226 & H$\beta$,[O{\sc iii}],H$\alpha$,[N{\sc ii}] & 9.0650 & SF \\
		SF & 99.9 & 99.9 \\
		            \hline
	\end{tabular}
	\end{table*}

\bsp	
\label{lastpage}
\end{document}